\documentclass[english]{article}
\usepackage[T1]{fontenc}
\usepackage[latin9]{inputenc}
\usepackage{geometry}
\usepackage{color}
\usepackage{amsmath}
\usepackage{graphicx}
\usepackage{setspace}
\usepackage[authoryear]{natbib}
\makeatletter

\providecommand{\tabularnewline}{\\}

\makeatother

\usepackage{babel}
\begin{document}

\title{Estimating the Competitive Storage Model: A Simulated Likelihood
Approach\thanks{The authors are indebted to the Editor; Ana Colubi, two anonymous
reviewers, Frank Asche, Hans Karlsen, Hans Julius Skaug and Bård Støve
for comments that greatly improved the paper.}}

\author{Tore Selland Kleppe\thanks{University of Stavanger, Department of Mathematics and Natural Sciences
(Corresponding author: email: tore.kleppe@uis.no, address: University
of Stavanger, 4036 Stavanger, Norway, telephone: +47 51831717, fax:
+47 51831750)}\and Atle Oglend\thanks{University of Stavanger, Department of Industrial Economics }}
\maketitle
\begin{abstract}
This paper develops a particle filter maximum likelihood estimator
for the competitive storage model. The estimator is suitable for inference
problems in commodity markets where only reliable price data is available
for estimation, and shocks are temporally dependent. The estimator
efficiently utilizes the information present in the conditional distribution
of prices when shocks are not iid. Compared to Deaton and Laroque's
composite quasi-maximum likelihood estimator, simulation experiments
and real-data estimation show substantial improvements in both bias
and precision. Simulation experiments also show that the precision
of the particle filter estimator improves faster than for composite
quasi-maximum likelihood with more price data. To demonstrate the
estimator and its relevance to actual data, we fit the storage model
to data set of monthly natural gas prices. It is shown that the storage
model estimated with the particle filter estimator beats, in terms
of log-likelihood, commonly used reduced form time-series models such
as the linear AR(1), AR(1)-GARCH(1,1) and Markov Switching AR(1) models
for this data set. 
\end{abstract}


\newpage{}

\section{Introduction}

This paper addresses the problem of estimating the structural parameters
of the competitive storage model when only price data is available
for estimation and supply shocks are temporally dependent. We propose
and investigate a particle filter estimator based on recently developed
methods in the particle filter literature \citep{gordon_salmond_smith_93,fv_rr_2007,Malik2011,DeJong01042013}.
We demonstrate that this estimator has superior large sample properties
and improved parameter identification properties over the conventional
composite pseudo maximum likelihood estimator (CML) \citep{Deaton1995,Deaton1996}.
Compared to the CML, the estimator also displays substantial reduction
in bias when it comes to various predicted price characteristics,
including price autocorrelation, where the CML estimator appears to
underestimate price persistence.

Particle filter methods are suitable to address inference problems
in non-linear models with latent variables, and appear especially
relevant for low dimensional models such as the partial equilibrium
competitive storage model in this paper. Our proposed estimator has
the added benefit of being an extension of the CML approach. The CML
estimator utilizes invariant distributions to integrate out the latent
state variable, in this case the supply shock. This procedure neglects
important information present in the conditional price distribution
when shocks are temporally dependent. It is this information the particle
filter estimator makes use of when inferring structural parameters. 

Improving structural model estimates provides better comparison between
competing commodity price models. We apply our estimator to a natural
gas price data set, and compare the structural model to popular reduced
form models such as the linear AR(1) model, AR(1)-GARCH(1,1) model
and a Markov Switching AR(1) model. The comparison to reduced form
models allows identification of what price features the storage model
is able to account for, and where improvements are needed. For the
natural gas market, the storage model performs better than all the
reduced form time-series models considered, in terms of log-likelihood.
This suggests that the non-linearity in the model, arising from the
non-negativity constraint on storage, is relevant for natural gas
price characteristics. As support for the relevance of the storage
model for natural gas we also find a relatively strong correspondence
between observed and model implied storage. The overall diagnostics
show that the storage model addresses features in the higher moments
of prices, specifically linked to excess kurtosis and ARCH effects. 

The competitive storage model is the prevailing economic model explaining
price dynamics of storable commodities. The model incorporates effects
of speculative storage behavior on market dynamics in a rational expectations
framework. The model in its current form can trace its origins to
the Theory of Storage (\citealp{williams1936speculation,Kaldor1939a,Working1949}).
Since the storage model explicitly recognizes the role of profit maximizing
storage behavior, it is often used to examine implications of different
commodity market policies and regulations. \citet{miranda1988effects}
used the storage model to investigate the effects of government price
support programs in the U.S. soybean market. \citet{gouel2013optimal}
analyzed various food price stabilization policies (public stocks,
state contingent subsidy/tax on production) using an extension of
the competitive storage model with risk averse consumers and incomplete
insurance markets. Similarly, \citet{brennan2003price} used the storage
model to analyze the effects of public interventions in the Bangladesh
rice market. Meaningful policy evaluations using structural models
require suitable model parameters. 

The theoretical foundation of the model is well established by now.
What has proved more difficult is confronting the model with data
in order to operationalize it for practical policy analysis and hypothesis
testing. Estimation has been limited by two factors: (1) the lack
of a closed form solution to the model, necessitating a computationally
demanding subroutine to solve a dynamic optimization problem for each
parameter evaluation, and (2) lack of reliable quantity data. With
improved computational power the first factor is becoming less important,
although the curse of dimensionality quickly rebalances growth in
computational power. For data availability, some commodity markets
do have reliable quantity data. The London Metal Exchange (LME), for
instance, provides information on metal stocks in LME warehouses (see
\citet{geman2013theory} for an analysis of the Theory of Storage
as applied to the LME metals). \citet{miranda1993estimation} estimate
the storage model using stock and price data from the U.S. soybean
market. Price data however remains the most accessible, and arguably
highest quality, commodity data. Because of this, price data should
in general be used when estimating the storage model. We provide simulation
evidence which shows that price data is sufficient to fully identify
the structural parameters of the model when the particle filter estimator
is applied, even when shocks are not iid. For model verification and
hypothesis testing it is also desirable to retain some data to test
model predictions. We illustrate this in the empirical application
part of the paper when we compare model implied storage, derived using
only price data, with actual storage. 

An early step towards empirical testing of the model can be found
in \citet{Wright1982}. The authors provide numerical solutions to
the model under different specifications, and confirm that model implied
behavior is consistent with observations of some important commodities.
One important feature of the storage model is the non-negativity constraint
on storage. This constraint makes the model non-linear; the no-arbitrage
restriction characterizing profit maximizing storage can fail if the
commodity is sufficiently scarce. Prices will move between two pricing
regimes defined by whether or not the no-arbitrage restriction holds.
Which regime is active depends on current price relative to a threshold
price marking the cut-off point where the no-arbitrage restriction
fails. \citet{Deaton1992} and \citet{Ng1996} estimate threshold
prices for various commodities using a generalized method of moments
estimator and find evidence for regime behavior consistent with model
predictions. 

Full structural estimation without quantity data was achieved in a
series of pioneering papers by \citet{Deaton1992,Deaton1995,Deaton1996}.
Deaton and Laroque estimated the model on price data by means of a
composite pseudo maximum likelihood estimator. Although a significant
step forward, the efficiency and precision of the estimator has been
questioned. \citet{Michaelides2000} show, using Monte Carlo simulations,
that the CML estimator tends to bias the estimates. The authors also
find that none of the simulation estimators beat the CML estimator
in a mean-squared sense, but that they have improved bias properties.
The particle filter estimator in this paper addresses the CML estimator
bias. \citet{cafiero2006storage} argue that the crucial kink in the
pricing function, due to the non-negativity constraint on storage,
is imprecisely estimated because of the smoothing effects of spline
methods used to interpolate between grid points. \citet{cafiero2011empirical}
show that precision can be greatly improved by increasing the number
of grid points used to approximate the pricing function. Recently
\citet{Cafiero01012015} proposed a maximum likelihood estimator of
the storage model with small properties superior to that of the pseudo
maximum likelihood approach. Their estimator relies of iid normal
supply shocks, and as such does not allow for temporal shock dependence.
The estimator considered in this paper provides an alternative to
the maximum likelihood estimator of \citet{Cafiero01012015} in the
case of temporal shock dependence.

By allowing for shock dependence, our paper deviates somewhat from
the empirical literature on the model. With the exception of \citet{Deaton1995,Deaton1996},
most applications assume iid shocks. One problem iid shocks is that
predicted price persistence is lower than what is typically observed.
This can be partly remedied by increasing the number of grid points
used to approximate the pricing function \citep{cafiero2011empirical}.
Extensions to the basic model such as allowing for an explicit convenience
yield component \citep{miranda1996empirical,ng2000explaining}, gestation
lags in production \citep{ng2000explaining}, the effect of news on
future production \citep{osborne2004market} and storage frictions
\citep{mitraille2009monopoly} have also been shown to increase model
price autocorrelation. The basic model has also been extended to a
more general equilibrium framework \citep{funke2011reviving,arseneau2013commodity},
and to consider monopolistic behavior in speculative storage \citep{mitraille2009monopoly}.
Temporal dependence in supply and/or demand shocks is not unreasonable
for the types of partial equilibrium models that the basic competitive
storage model represents. Any non-modeled stochastic exogenous effects
must enter through the specified shock processes. Slowly changing
macroeconomic conditions, for instance, is likely to give rise to
persistence in demand for the commodity. Overall, it seems unreasonable
to expect that speculative storage alone can fully account for the
substantial shock persistence observed in commodity prices. In the
additive, single shock, formulation of the standard competitive storage
model, positive demand shocks are equivalent to negative supply shocks,
and supply shocks should be interpreted as net-supply shocks. 

Another issue in the Deaton and Laroque CML estimation is the use
of observed Fisher information to derive asymptotic standard deviations
of estimated parameters. However, we have found that asymptotic results
can be misleading for sample sizes relevant to this methodology. To
work around this complication, we make no effort to make the log-likelihood
smooth. Rather we introduce a full parametric specification consistent
with the moments of the storage model and rely on parametric bootstrap
for estimating statistical standard errors. The validity of the parametric
specification is then tested via generalized residuals.

The paper proceeds as follows. In section 2 we give a description
of the storage model. Section 3 describes the estimation methodology.
Following this we investigate the performance of the estimator on
simulated data in section 4. Comparing the estimator to the CML estimator
we find that bias is substantially reduced. In addition the precision
of key structural parameters are greatly improved. In section 5 we
apply the estimation procedure to a natural gas data set and find
evidence that the non-negativity constraint on storage is relevant
in describing key properties of prices. Finally, section 6 concludes.

\section{The Storage Model}

Before turning to the issue of estimating structural parameters, we
provide a brief description of the basic competitive storage model
with autocorrelated supply shocks. The model is the same as in \citet{Deaton1995,Deaton1996},
from now on referred to as DL. For more details on the model see \citet{Deaton1996}.

Assume at any time there is exogenous stochastic supply $z_{t}$,
which follows a first-order linear autoregressive process: $z_{t}=\rho z_{t-1}+\epsilon_{t}$,
where $\epsilon_{t}$ is a standard normal random variable. The supply
shock is what fundamentally drives variations in prices. The market
consists of consumers and risk-neutral competitive speculators that
hold inventories. Following DL, we assume a proportional decay of
stocks in storage. The constant depreciation rate $\delta$ accounts
for the direct cost of storage, and is a structural parameter to be
estimated. Let $I_{t}$ be the level of inventories at time $t$.
The amount of stocks on hand $x_{t}$ and supply $z_{t}$ then follow
the laws of motion:

\begin{equation}
x_{t}=\left(1-\delta\right)I_{t-1}+z_{t},\label{eq:X_dymamics}
\end{equation}

\begin{equation}
z_{t}=\rho z_{t-1}+\epsilon_{t}.\label{eq:Z_dymamics}
\end{equation}

The problem facing speculators is choosing the level of inventories
to maximize the expected discounted profits from storage. In this
paper, as in DL, consumers are assumed to hold linear inverse demand,
represented by $P\left(z\right)=a+bz$, where $a$ and $b<0$ are
structural parameters. The opportunity cost of capital tied to storage
is assumed a fixed real interest rate $r$ per period. Combined with
the depreciation rate, $\beta=\frac{1-\delta}{1+r}$ accounts for
the cost of storing one unit of the commodity for one period. \textcolor{black}{There
are two state variables relevant to the optimal storage decision at
any time, the current stock level $x_{t}$ and the current state of
the supply shock $z_{t}$.} Let $V\left(x_{t},z_{t}\right)$ be the
value of the commodity stock at time $t$ given competitive speculators
follow an optimal storage policy. This value function must satisfy
the Bellman functional equation 

\begin{equation}
V\left(x_{t},z_{t}\right)=\underset{I_{t}}{\max}\left\{ p_{t}\left(x_{t}-I_{t}\right)+\beta E_{t}V\left(\left(1-\delta\right)I_{t}+z_{t+1},\rho z_{t}+\epsilon_{t+1}.\right)\right\} ,\label{eq:V_func}
\end{equation}

where the maximization is subject to equations (\ref{eq:X_dymamics}),(\ref{eq:Z_dymamics})
and the non-negativity constraint $I_{t}\geq0$. Note that $x_{t}-I_{t}$
is the amount supplied to the market, and $p_{t}$ the commodity price
at time $t$. The price is considered fixed when making storage decisions
and speculators are assumed to hold rational expectations. Given storage
is not bounded, $I_{t}>0$, profit maximizing competitive storage
implies the no-arbitrage restriction: $\beta E_{t}p_{t+1}=p_{t}$
(this follows from taking the derivative of equation \ref{eq:V_func}
w.r.t. to $I_{t},$ setting equal to zero, and applying the envelope
theorem). The no-arbitrage relationship will fail if $P\left(x_{t}\right)\geq\beta E_{t}p_{t+1}$,
in which case optimal storage must be $I_{t}=0$ (stock-out). The
optimal storage policy implies the following restriction:

\begin{equation}
p_{t}=\max\left[P\left(x_{t}\right),\beta E_{t}p_{t+1}\right]\label{eq:arb_rel}
\end{equation}

The restriction states that if selling the entire stock $x_{t}$ gives
a higher price than what could be gained from storing, zero storage
is optimal and current price is dictated by consumer willingness to
pay for existing stocks, $P\left(x_{t}\right)$. Otherwise, the no-arbitrage
condition holds and profit maximizing storage ensures current price
$p_{t}$ equal to $\beta E_{t}p_{t+1}$. 

Provided supply has compact support and inventories are costly $\beta<1$,
\citet{Deaton1992} and \citet{Chambers1996} prove that a solution
to (\ref{eq:arb_rel}) exist in the form of a rational expectations
pricing function $p=f\left(x,z\right)$. The pricing function $f\left(x,z\right)$
gives the competitive equilibrium price $p$ consistent with optimal
storage at states $x$ and $z$ and market clearing. The function
in general has no closed form solution and must be solved for numerically.
The mapping $p=f\left(x,z\right)$ is the solution to the functional
equation:

\begin{equation}
f\left(x,z\right)=\max\left[\beta\int f\left(\epsilon+\rho z+\left(1-\delta\right)\left(x-P^{-1}\left(f\left(x,z\right)\right)\right),\epsilon+\rho z\right)d\Phi\left(\epsilon\right),P\left(x\right)\right]\label{eq:f_fixedpoint}
\end{equation}
where $\Phi\left(\epsilon\right)$ is the standard normal distribution
function. For given parameters, the function can be found by conventional
numerical procedures, we refer to section 3.1 for the specific numerical
procedure used in this paper. The price function provides the means
to generating the predictive moments necessary for estimation purposes. 

\section{Estimation Methodology}

The problem at hand is to estimate the structural parameters: $\theta=\left[\rho,a,b,\delta\right]$.
By utilizing the price function (\ref{eq:f_fixedpoint}) we can construct
the one-period ahead mean and variance of price as a function of current
price $p_{t}$ and supply shock $z_{t}$. We define these quantities
respectively as $\mu(p_{t},z_{t}):=E(p_{t+1}|p_{t},z_{t})$ and $\sigma^{2}(p_{t},z_{t}):=Var(p_{t+1}|p_{t},z_{t})$.
This procedure allows the construction of the likelihood over all
observed prices. For the model with autocorrelated shocks, the conditional
mean and variance depend on the unobserved supply shock $z_{t}$.
As such $z_{t}$ must be integrated out of the conditional moments.
To achieve this, DL integrate over the invariant distributions of
$z_{t}$ conditionally on $p_{t}$. However, using invariant distributions
do not utilize the full information available from the observables
about the state of the system, and is likely to be inefficient, as
was recognized by DL themselves. Simulation evidence provided below
also suggest that the invariant distribution integration does not
lead to improved large sample properties of the estimator; the bias
of the estimates does not appear to shrink as more price data becomes
available. In the following we propose a particle filter approach
to deal with this problem. The particle filter avoids using invariant
distributions, which provides more efficient and precise estimates. 

Throughout we let $\mathcal{N}(x;\mu,\sigma^{2})$ denote the $N(\mu,\sigma^{2})$
density evaluated at $x$. Moreover, for consistency with the notation
of DL who already use $p$ and $f$, we use $\pi(x)$ as the generic
symbol for probability density functions. 

\subsection{Numerical solution of the price function equation}

As for most non-linear rational expectation models, obtaining a closed
form solution to the dynamic optimization problem characterized by
(\ref{eq:f_fixedpoint}) is not possible, and we must resort to numerical
methods. A large body of literature, surveyed in e.g. \citet{miranda1997,Aruoba20062477,gouel2013comparing},
is devoted to such numerical solutions. The price function $f(x,z)$
is known to have discontinuous derivatives w.r.t. $x$ at the point
$x^{*}(z)$, the threshold stock level where the no-arbitrage restriction
breaks down for stocks below this level, i.e. where $I=0$ for $x\leq x^{*}(z)$.
The threshold stock level is characterized by 
\[
x^{*}(z)=\sup\{x;\;f(x,z)=P(x)\}.
\]
To solve for the price function $f\left(x,z\right)$ we iterate on
the recursive formulation of equation \ref{eq:f_fixedpoint}. This
provides a quick and reliable numerical approximation to $f\left(x,z\right)$.
Perturbation- and spectral methods appear less applicable for non-smooth
solutions, and are also in general more time consuming. Our implementation
is similar in spirit to that of DL, but differs in particular in the
choices of grids. 

\subsubsection{Price function iteration grid}

When choosing the grid over which the price function is computed,
several aspects have to be taken into consideration. Firstly, the
grid has to be fine enough to avoid substantial biases in the parameter
estimates \citep{cafiero2011empirical}. Moreover, since the price
function is computed for many sets of parameters during the numerical
maximization of the likelihood, computational tractability must be
kept in mind. Finally, the price function must be continuous w.r.t.
the parameters, and therefore rules out any adaptive strategies for
finding grids.

The specific grids employed are in the $z$-direction (supply shocks)
equally spaced with $M_{z}$ grid-points $Z_{1},\dots,Z_{M_{z}}$.
Throughout most of this work, $M_{z}$ is set to 64. The grid is chosen
to cover 6 standard deviations of the unconditional distribution of
$z_{t}$, corresponding to $Z_{1}=-6/\sqrt{1-\rho^{2}}$ and $Z_{M_{z}}=6/\sqrt{1-\rho^{2}}$.
Compared to DL, who used 11 grid points and a discretization similar
to that of \citet{Tauchen1986177}, our grid is more finely spaced
and covers a larger range. This reflect both access to faster- and
parallel computing and the fact that the later described particle
filter sometimes request evaluations of $z\mapsto f(x,z)$ in areas
that are very unlikely to be visited under the marginal distribution
of $z_{t}$.

In the $x$-direction (stocks), we divide the grid over which $f(x,z)$
is computed into two parts. First, a finer equally spaced grid with
$M_{x,1}$ grid points covering the lower range of $x$ known to contain
$x^{*}(z)$ for all relevant values of $z$. Second, a coarser equally
spaced grid with $M_{x,2}$ grid points covering the higher range
of $x$ where $x\mapsto f(x,z)$ is slowly varying. The grid points
are denoted by $X_{1},\dots,X_{M_{x,1}+M_{x,2}}$, and for most of
this work, $M_{x,1}=M_{x,2}=128$. The range of the first grid is
set to 
\[
X_{1}=\min\left(P^{-1}(p_{max}),\;Z_{1}\right),\;X_{M_{x,1}}=\max\left(-\frac{a}{b},\;Z_{M_{z}}\right).
\]
where $p_{max}$ is larger than the observed maximum price\footnote{In this work we typically set $p_{max}=20$ as the real data sets
are normalized to have unit mean.}. In the definition of $X_{1}$, $P^{-1}(p_{max})$ ensures that the
numerical representation of $f(x,z)$ spans sufficiently high prices,
and $Z_{1}$ reflects that stock at hand cannot be smaller than the
smallest supply \citep{Deaton1996}. In the upper limit of the first
grid, $-a/b$ ensures that $X_{M_{x,1}}\geq x^{*}(z)$ since $f(-a/b,z)>P(-a/b)=0$.
Moreover, $Z_{M_{z}}$ is chosen less rigorously based on experience
to ensure numerical accuracy in cases when the demand slope $b$ is
large in magnitude. The second grid is uniformly spaced with
\[
X_{M_{x,1}+1}=X_{M_{x,1}}+(X_{M_{x,1}+M_{x,2}}-X_{M_{x,1}})/M_{x,2},\;X_{M_{x,1}+M_{x,2}}=cZ_{M_{z}}/\delta.
\]
The upper $x$-range of the grid is inspired by DL, who obtain that
the stock at hand $x_{t}$ asymptotes to $Z_{M_{z}}/\delta$ when
the supply is always at maximum $Z_{M_{z}}$ and no stock is consumed.
However, we have included the factor $c=1.5$ obtained by trial and
error to ensure that sufficiently low prices are always represented,
in particular when the numerical optimizer tries values of the parameters
that are inconsistent with the data at hand.

\subsubsection{Price function iteration}

Equation (\ref{eq:f_fixedpoint}) defines a functional fixed point
for $f(x,z)$ which we iterate from an initial guess. Our numerical
solution, say $\hat{f}(x,z)$, obtains as a bilinear interpolation
between tabulated values over the previously discussed grid. The integral
in (\ref{eq:f_fixedpoint}) for each grid point $(X_{i},\;Z_{j})$
is approximated as 
\begin{eqnarray}
 &  & \int f\left(\epsilon+\rho Z_{j}+\left(1-\delta\right)\left(X_{i}-P^{-1}\left(f\left(X_{i},Z_{j}\right)\right)\right),\epsilon+\rho Z_{j}\right)d\Phi(\epsilon),\label{eq:vfi_integral_true}\\
 & \approx & \int\hat{f}\left(\eta+\left(1-\delta\right)\left(X_{i}-P^{-1}\left(\hat{f}\left(X_{i},Z_{j}\right)\right)\right),\eta\right)\mathcal{N}(\eta;\rho Z_{j},1)d\eta,\label{eq:vfi_integral_intermediate}\\
 & \approx & \sum_{k=1}^{M_{z}}\hat{f}(Z_{k}+\left(1-\delta\right)\left(X_{i}-P^{-1}\left(\hat{f}\left(X_{i},Z_{j}\right)\right)\right),Z_{k})W_{j,k},\label{eq:vfi_integral_approximation}
\end{eqnarray}
where the weight matrix $W$ has elements 

\[
W_{j,k}=\frac{\mathcal{N}(Z_{k};\rho Z_{j},1)}{\sum_{l=1}^{M_{z}}\mathcal{N}(Z_{l};\rho Z_{j},1)},\;j,k=1,\dots,M_{z}.
\]
I.e. each row in the weight matrix represents discrete distribution
over $\{Z_{1},\dots,Z_{M_{z}}\}$ having probability masses proportional
to $\mathcal{N}(Z_{k};\rho Z_{j},1),\;k=1,\dots,M_{z}$. The reason
for choosing this particular discretization follows the insight of
DL that it leads to $\hat{f}(x,z)$ being evaluated only at grid points
in the $z$-direction. Thus the evaluation of each term in the sum
(\ref{eq:vfi_integral_approximation}) is reduced to a univariate
linear interpolation problem in the $x$-direction. 

Equipped with the integral approximation (\ref{eq:vfi_integral_true}
- \ref{eq:vfi_integral_approximation}), the price function iteration
is started at $\hat{f}(x,z)=\max(P(x),0)$, and proceeds by alternating
between 
\begin{eqnarray*}
G(X_{i},Z_{j}) & = & \beta\sum_{k=1}^{M_{z}}\hat{f}(Z_{k}+\left(1-\delta\right)\left(X_{i}-P^{-1}\left(\hat{f}\left(X_{i},Z_{j}\right)\right)\right),Z_{k})W_{j,k},\\
\hat{f}(X_{i},Z_{j}) & = & \max(P(X_{i}),G(X_{i},Z_{j})),
\end{eqnarray*}
for $i=1,\dots,M_{x,1}+M_{x,2},\;j=1,\dots,M_{z}$. Throughout this
work, we perform 400 iterations, as a fixed number of iterations are
required to obtain a continuous likelihood. This number of iterations
is typically sufficient to bring maximal absolute change in $\hat{f}$
at the last iteration to $\sim1e-3$. The outer loop over $i$ is
performed in parallel over 8 kernels on the computer applied and the
overall routine for calculating $\hat{f}$ requires $\sim1$ second
when implemented in Fortran 90.

\subsubsection{Predictive moments}

The predictive moments $\mu(p_{t},z_{t})=E(p_{t+1}|p_{t},z_{t})$
and $\sigma^{2}(p_{t},z_{t})=Var(p_{t+1}|p_{t},z_{t})$ are required
to form the likelihood function of the data. We compute them in two
steps. First, the time $t$ stock at hand $x_{t}=x_{t}(p_{t},z_{t})$
is recovered, modulo price function approximation error, from $p_{t}$
and $z_{t}$ by solving $p_{t}=\hat{f}(x_{t},z_{t})$ for $x_{t}$.
In practice, this is implemented using a binary search in the $x$-direction.
As bilinear interpolation is employed to compute $\hat{f}(x,z)$ off
the grid, we have not encountered cases where $x\mapsto\hat{f}(x,z)$
is not monotone, and therefore no problems with non-uniqueness of
$x_{t}$ was encountered. 

Secondly, the predictive mean is calculated using that
\begin{eqnarray*}
\mu(p_{t},z_{t}) & = & E(p_{t+1}|p_{t},z_{t})=E(f(x_{t+1},z_{t+1})|p_{t},z_{t})\\
 & = & E\left(f\left((1-\delta)I_{t}+z_{t+1},z_{t+1}\right)|p_{t},z_{t}\right),\\
 & = & \int f\left((1-\delta\right)\left[x_{t}(p_{t},z_{t})-P^{-1}(p_{t})\right]+\rho z_{t}+\epsilon_{t+1},\rho z_{t}+\epsilon_{t+1})d\Phi(\epsilon_{t+1}),
\end{eqnarray*}
and a completely analogous argument applies for the conditional variance
$\sigma^{2}(p_{t},z_{t})$. Storage $I_{t}$ is here derived as the
difference between implied stock $x_{t}(p_{t},z_{t})$ and consumption
$P^{-1}(p_{t})$. In practice, we implement these integrals using
16-point Gauss-Hermite quadrature and substituting $\hat{f}$ for
$f$.

\subsection{Inference}

In their inference, DL do not specify a parametric family for $p_{t+1}|p_{t},z_{t}$,
and use quasi maximum likelihood\footnote{ In the sense of substituting unspecified families of distributions
with Gaussians \citep{gourieroux_quasi_theory}.} for estimation. Here we depart in taking $p_{t+1}|p_{t},z_{t}$ to
be Gaussian to obtain a, conditionally on $p_{1}$, complete parametric
model
\begin{eqnarray}
p_{t+1} & = & \mu(p_{t},z_{t})+\sqrt{\sigma^{2}(p_{t},z_{t})}\eta_{t+1},\;\eta_{t}\sim\text{i.i.d.}\;N(0,1),\;t=1,\dots,T-1,\label{eq:stoch_model1}\\
z_{t+1} & = & \rho z_{t}+\epsilon_{t+1},\;\epsilon_{t}\sim\text{i.i.d.}\;N(0,1),\;t=1,\dots,T-1,\label{eq:stoch_model2}\\
z_{1} & \sim & N\left(0,\frac{1}{1-\rho^{2}}\right),\label{eq:stoch_model3}
\end{eqnarray}
that is consistent with the moments of the price process of the storage
model. It is worth noticing that $p_{t+1}|p_{t},z_{t}\sim N(\mu(p_{t},z_{t}),\sigma^{2}(p_{t},z_{t}))$
does not imply that $p_{t+1}|p_{t}$ or $p_{t+1}|p_{1},\dots,p_{t}$
are Gaussian. Rather, the unconditional transition laws are complicated
mean-variance mixtures of Gaussians that can exhibit skewness and
heavy tails. 

The reason for choosing a parametric model is mainly for convenience,
as doing so grants us access to the likelihood analysis toolbox and
parametric bootstrapping. Moreover, the diagnostics reported later
allow us to test whether imposing a Gaussian distribution on $\eta_{t}$
is reasonable. Quasi maximum likelihood, on the other hand, would
be complicated due to the lack of reliable derivatives to form sandwich
formulas, and is not easily bootstrapped. 

The model (\ref{eq:stoch_model1}-\ref{eq:stoch_model3}) is, for
observations $p_{1},\dots,p_{T}$, a dynamic latent variable model,
with dependence structure typical of a time-discretized diffusion
model, and the conditional likelihood function is expressed in terms
of
\begin{equation}
L(\theta|p_{1},\dots,p_{T})=\pi(p_{2},\dots,p_{t}|p_{1})=\int\pi(z_{1})\prod_{t=2}^{T}\pi(p_{t},z_{t}|p_{t-1},z_{t-1})dz_{1}\cdots dz_{T}.\label{eq:marginal_likelihoood}
\end{equation}
To circumvent integrating over $z_{t}$, DL uses a composite likelihood
technique \citep{lind:1988} where the product of likelihoods of consecutive
pairs of prices is substituted for the full likelihood. Using composite
likelihoods is known to result in loss of estimation efficiency \citep{varin_vidoni_2008}
over full likelihood-based techniques as we propose here. A large
body of literature is devoted to fully likelihood-based inference
in models with this structure, and include Bayesian Markov chain Monte
Carlo (MCMC) \citep{eraker_2001} and simulated maximum likelihood
\citep{durbin_koopman_97,shepard_pitt97,durham06}. As the calculation
of $\hat{f}$ for each combination of the parameters is expensive,
MCMC (including the particle MCMC approach of \citet{RSSB:RSSB736,Pitt2012134})
would be very time consuming. Moreover, the fact that $z_{t}\mapsto\log\pi(p_{t+1}|p_{t},z_{t})$,
implied by (\ref{eq:stoch_model1}), is a non-smooth function bars
the usage of smoothing-based importance samplers such as \citet{durbin_koopman_97,shepard_pitt97,liesenfeld_richard_03,richardetal07}.
Consequently, we resort to simulated maximum likelihood, where the
likelihood function is calculated using a continuous particle filter
in the spirit of \citet{Malik2011} to obtain efficient estimators
and retain computational tractability.

\subsubsection{Simulated likelihood}

To estimate the marginal log-likelihood $l(\theta)=\log L(\theta|p_{1},\dots,p_{T})$
at a fixed parameter $\theta$, we take as vantage point the sampling
importance resampling (SIR) particle filter \citep{gordon_salmond_smith_93},
adapted to time-discretized diffusion structures as described in \citet{durham06}.
The filter with $N$ particles may be described as follows:
\begin{enumerate}
\item Initialization: set $t=1$, $\hat{l}=0$ 
\item Simulate $z_{t,pred}^{(j)}\sim\pi(z_{1}),\;j=1,\dots,N$.
\item Compute $w_{t}^{(j)}=\pi(p_{t+1}|p_{t},z_{t,pred}^{(j)})$ and set
$L_{t}=\frac{1}{N}\sum w_{t}^{(j)}$. {[}$L_{t}$ is an approximation
to $\pi(p_{t+1}|p_{1},\dots,p_{t})${]}
\item Set $\hat{l}\leftarrow\hat{l}+\log(L_{t})$. {[}$\hat{l}$ is an approximation
to $\log\pi(p_{2},\dots,p{}_{t+1}|p_{1})${]}
\item Re-sample $N$ particles from $z_{t,pred}^{(j)}$ with weights $w_{t}^{(j),*}=$$w_{t}^{(j)}/\sum_{k=1}^{N}w_{t}^{(k)}$
to form $z_{t,filt}^{(j)},\;j=1,\dots,N$. {[}$\left\{ z_{t,filt}^{(j)}\right\} _{j=1}^{N}$
is an equally weighted representation of $\pi(z_{t}|p_{1},\dots,p_{t+1})${]}
\item Simulate $z_{t+1,pred}^{(j)}\sim\pi(z_{t+1}|z_{t}=z_{t,filt}^{(j)}),\;j=1,\dots,N$.
{[}$\left\{ z_{t+1,pred}^{(j)}\right\} _{j=1}^{N}$ is an equally
weighted representation of $\pi(z_{t+1}|p_{1},\dots,p_{t+1})${]}
\item If $t<T-1,$ set $t\leftarrow t+1$ and go to step 3.
\end{enumerate}
It is well known \citep[see e.g.][]{Malik2011} that the resampling
in step 5 originate discontinuous likelihood approximations even if
the same random numbers are used for repeated evaluation of $\hat{l}(\theta)$
for different values of the parameters, and thereby renders subsequent
numerical maximization difficult. \citet{Malik2011} propose to obtain
draws in step 5 based on a smoothed version of the empirical distribution
function of the weighted sample $(z_{t,pred}^{(j)},w_{t}^{(j),*})$.
We follow a different path to obtain a continuous likelihood, in combining
step 5 and 6. Write 
\[
\pi(z_{t}|p_{1},\dots,p_{t+1})\approx\sum_{j=1}^{N}w_{t}^{(j),*}\delta(z_{t}-z_{t,pred}^{(j)})
\]
where $\delta(\cdot)$ denotes a unit point mass in $0$. Then 
\[
\pi(z_{t+1}|p_{1},\dots,p_{t+1})\approx\sum_{j=1}^{N}w_{t}^{(j),*}\mathcal{N}(z_{t+1}|\rho z_{t,pred}^{(j)},1)
\]
which can be sampled from continuously using inversion sampling. We
use a fast Fourier transform method \citep{kleppe_skaug_pspf} based
on stratified uniform common numbers described in Appendix A which
has the same computational complexity of $O(N\log_{2}(N))$ as the
method of \citet{Malik2011}. Notice that advancing to the supply
process (step 6) does not involve simulation, which will have desirable
effect on Monte Carlo variability of overall Monte Carlo estimate
of the log likelihood $\hat{l}(\theta)$. 

A number of improved particle filters have been proposed in literature
\citep[see][for a survey]{cappe_godsill_moulines_2007} with the prospect
of further reducing Monte Carlo variation. However, for relevant ranges
of the parameters, the model (\ref{eq:stoch_model1}-\ref{eq:stoch_model3})
has a low signal-to-noise ratio. Thus does our simple modified SIR
filter produce sufficiently precise estimates of the log-likelihood
for moderate $N$. E.g. for $T\sim1000$, $N=4096$ and the filter
implemented in Fortran 90 produces acceptable accuracy while the computing
times of an evaluation of $\hat{l}(\theta)$ is on the same order
as the time required to compute $\hat{f}$.

\subsubsection{Algorithm, estimation and standard errors}

Based on the above introduced notation, to evaluate $\hat{l}(\theta)$
for any given $\theta$ and price series $p_{1},\dots,p_{T}$ requires
the following steps:
\begin{enumerate}
\item Solve numerical the price function $\hat{f}(x,z)$ as described in
section 3.1-3.1.2 for parameter $\theta$.
\item Set $t=1$, $\hat{l}=0$ and simulate $z_{t}^{(j)}\sim N(0,1/(1-\rho^{2})),\;j=1,\dots,N$.
\item Compute $\mu(p_{t},z_{t}^{(j)}),$ $\sigma^{2}(p_{t},z_{t}^{(j)})$
for $j=1,\dots,N$ based on the numerical solution of the price function
as described in Section 3.1.3 and set $w_{t}^{(j)}=\mathcal{N}(p_{t+1}|\mu(p_{t},z_{t}^{(j)}),\sigma^{2}(p_{t},z_{t}^{(j)}))$,
$w_{t}^{(j),*}=w_{t}^{(j)}/\sum_{k=1}^{N}w_{t}^{(k)}$ for $j=1,\dots,N$.
\item Set $L_{t}=\frac{1}{N}\sum w_{t}^{(j)}$ and $\hat{l}\leftarrow\hat{l}+\log(L_{t})$. 
\item Simulate $z_{t+1}^{(j)}\sim\sum_{k=1}^{N}w_{t}^{(k),*}\mathcal{N}(z_{t+1}|\rho z_{t}^{(k)},1),\;j=1,\dots,N$
using the FFT-based method described in Appendix A.
\item If $t<T-1,$ set $t\leftarrow t+1$ and go to step 3. Otherwise return
$\hat{l}(\theta)=\hat{l}$.
\end{enumerate}
The use of bilinear interpolation for evaluation of $\hat{f}$ off
the grid results in $\hat{l}(\theta)$ having discontinuous derivatives.
Thus gradient-based optimizers are not suitable, and we therefore
find the simulated maximum likelihood estimator $\hat{\theta}=\arg\max\hat{l}(\theta)$
using the Nelder-Mead type optimizer implemented in MATLAB's $\mathtt{fminsearch}$
routine. 

The lack of continuous derivatives also bars estimating reliable parameter
standard errors using the observed Fisher information matrix. Instead,
we rely on parametric bootstrap, where the simulated maximum likelihood
estimator is applied to synthetic data simulated from the model (\ref{eq:stoch_model1}-\ref{eq:stoch_model3}).

\subsubsection{Properties of the simulated maximum likelihood estimator}

All structural parameters in the model are statistically identified
given the non-negativity constraint on storage is not at all times
binding. The price process implied by the model moves between two
regimes dependent on the storage constraint. The only parameter unique
to the positive storage regime is depreciation $\delta$. As this
tends to one, storage becomes increasingly costly and the storage
regime less relevant. We show in the empirical application below that
under zero speculative storage, the price process reduces to a linear
AR(1) process, which is stationary and mean reverting given the supply
shock is stationary and mean reverting. At the other extreme, where
depreciation tends to $-r$, storage becomes costless and the storage
constraint less likely to be binding. The parameters will still be
identified, but the non-linearity arising from the regime-shifts will
be less relevant to the price dynamics. Indeed, the primary feature
of the DL formulation of the storage model is that the market should
move between regimes where speculative storage occurs and where there
is a speculative storage stock-out. The relevance of this regime shift
feature can be checked by looking at the probability of model implied
stock levels moving below a level where speculative storage is profitable.
We do this in our empirical application below.

\subsubsection{Diagnostics}

When carrying out diagnostics tests on time series models, a standard
approach is to carry out a battery of tests on the residuals. For
the dynamic latent variable models, calculating and characterizing
residuals are complicated by the presence of the latent factor. For
model (\ref{eq:stoch_model1}-\ref{eq:stoch_model3}), both $p_{t+1}|p_{t}$
and $p_{t+1}|p_{1},\dots,p_{t}$ are complicated non-linear (in $p_{t}$
or $p_{1},\dots,p_{t}$) mean-variance mixtures of Gaussians without
closed form expressions. Consequently we resort to generalized residuals
as explained in \citet{durham06}. For a correctly specified model,
the probability $u_{t}$ of observing $p_{t}$ or smaller, conditionally
on $p_{1},\dots,p_{t-1}$ should be iid uniformly distributed. Fortunately,
these generalized residuals $u_{t}$ can easily be estimated from
particle filter output \citep{durham06} as 
\begin{equation}
u_{t}\approx\hat{u}_{t}=\frac{1}{N}\sum_{j=1}^{N}\Phi\left(p_{t}|\mu(p_{t-1},z_{t-1}^{(j)}),\sigma^{2}(p_{t-1},z_{t-1}^{(j)})\right),\;t=2,\dots,T.\label{eq:generalized_residuals}
\end{equation}
Throughout this work, we use for diagnostics purposes the transformed
generalized residuals $\hat{\eta}_{t}=\Phi^{-1}(\hat{u}_{t}),\;t=2,\dots,T$,
from now on referred to as the residuals. Under a correctly specified
model $\left\{ \hat{\eta}_{t}\right\} _{t=2}^{T}$ should be an iid
sequence of standard Gaussian variables. However it is worth noticing
that $\left\{ \hat{\eta}_{t}\right\} _{t=2}^{T}$ being Gaussian does
not imply that $p_{t+1}|p_{1},\dots,p_{t},\;t=1,\dots,T-1$ are Gaussian,
only that the distributions of $p_{t+1}|p_{1},\dots,p_{t},\;t=1,\dots,T-1$
are correctly captured by the model.

\subsection{Composite likelihood estimator}

As a benchmark for our proposed simulated maximum likelihood estimator,
we also implement a composite quasi maximum likelihood estimator in
the spirit of DL. We use the same numerical solution procedure for
the price function $\hat{f}$. This method relies on looking at consecutive
pairs of prices $(p_{t},p_{t+1}),\;t=1,\dots,T-1$ and finding the
unconditional (with respect to $z_{t}$) versions of the predictive
moments $\mu$, $\sigma^{2}$ as
\begin{eqnarray}
\mu(p_{t}) & = & \int\mu(p_{t},z_{t})\pi(z_{t}|p_{t})dz_{t},\;t=1,\dots,T-1,\label{eq:CML_mean}\\
\sigma^{2}(p_{t}) & = & \int\sigma^{2}(p_{t},z_{t})\pi(z_{t}|p_{t})dz_{t},\;t=1,\dots,T-1,\label{eq:CML_var}
\end{eqnarray}
where the conditional $\pi(z_{t}|p_{t})$ derives from the joint stationary
distribution $\pi(p_{t},z_{t})$. For this purpose, DL derive the
full joint stationary distribution of $(x_{t},z_{t})$ on the grid
used for solving for $f$, and thereafter derive distributions of
$z_{t}$ conditional on each observed $p_{t}$. In our setup, we advise
against this practice, as it involves factorizing a $(M_{x,1}+M_{x,2})M_{z}\times(M_{x,1}+M_{x,2})M_{z}=16384\times16384$
unstructured matrix. Instead we rely on a Monte Carlo estimator of
$\pi(p_{t}|z_{t})$ which is implemented using the following steps:
\begin{enumerate}
\item Simulate a realization $\left\{ (\check{p}_{t},\check{z}_{t})\right\} _{t=1}^{n_{i}n_{t}}$
of the joint $(p_{t},z_{t})$ process (\ref{eq:stoch_model1}-\ref{eq:stoch_model3})
of length $n_{i}n_{t}.$
\item Subsample every $n_{t}$-th observation of the simulated process to
form the less autocorrelated sample $\tilde{S}=\left\{ \left(\tilde{p}_{i},\tilde{z}_{i}\right)\right\} _{i=1}^{n_{i}}$
where $\tilde{p}_{i}=\check{p}_{n_{t}(i-1)+1}$, $\tilde{z}_{i}=\check{z}_{n_{t}(i-1)+1}$. 
\item Calculate observed mean and variance of $\tilde{S}$, i.e. $m_{p}=\frac{1}{n_{i}}\sum_{i=1}^{n_{i}}\tilde{p}_{i}$,
$s_{p}^{2}=\frac{1}{n_{i}-1}\sum_{i=1}^{n_{i}}(\tilde{p}_{i}-m_{p})^{2}$,
and correspondingly for $m_{z},$ $s_{z}^{2}$.
\item Set up a uniform grid $\underline{Z}=\{\underline{z}^{(j)}\}_{j=1}^{n_{g}}$
covering $m_{z}\pm4\sqrt{s_{z}^{2}}.$
\item For each $t=1,\dots,T-1$; calculate un-normalized kernel estimate
with Gaussian kernels to $\pi(z_{t}|p_{t})$ over $\underline{Z}$
given by 
\[
\underline{w}_{t}^{(j)}=\sum_{i=1}^{n_{i}}\exp\left(-\frac{1}{2}\frac{(p_{t}-\tilde{p}_{i})^{2}}{h_{p}^{2}}-\frac{1}{2}\frac{(\underline{z}^{(j)}-\tilde{z}_{i})^{2}}{h_{z}^{2}}\right).
\]
Here the bandwidths are chosen via the simple plug-in rules $h_{p}=2n_{i}^{-\frac{1}{6}}\sqrt{s_{p}^{2}}$,
$h_{z}=2n_{i}^{-\frac{1}{6}}\sqrt{s_{z}^{2}}$.
\item For each $t=1,\dots,T-1$; approximate (\ref{eq:CML_mean}, \ref{eq:CML_var})
as 
\begin{eqnarray*}
\mu(p_{t}) & \approx & \bar{\mu}(p_{t})=\frac{\sum_{j=1}^{n_{g}}\underline{w}_{t}^{(j)}\mu(p_{t},\underline{z}^{(j)})}{\sum_{j=1}^{n_{g}}\underline{w}_{t}^{(j)}},\\
\sigma^{2}(p_{t}) & \approx & \bar{\sigma}^{2}(p_{t})=\frac{\sum_{j=1}^{n_{g}}\underline{w}_{t}^{(j)}\sigma^{2}(p_{t},\underline{z}^{(j)})}{\sum_{j=1}^{n_{g}}\underline{w}_{t}^{(j)}}.
\end{eqnarray*}
\end{enumerate}
The composite quasi log-likelihood function then becomes
\[
\bar{l}(\theta)=\sum_{t=1}^{T-1}\left(-\frac{\left[p_{t+1}-\bar{\mu}(p_{t})\right]^{2}}{2\bar{\sigma}^{2}(p_{t})}-\frac{1}{2}\log\left[2\pi\bar{\sigma}^{2}(p_{t})\right]\right),
\]
and the corresponding composite quasi maximum likelihood estimator
is given as $\bar{\theta}=\arg\max_{\theta}\bar{l}(\theta)$. In practice
we choose the tuning parameters to be $n_{i}=50,000$, $n_{t}=32$
and $n_{g}=128$, and therefore this routine is considerably more
expensive than the above described simulated maximum likelihood routine,
but still less costly than computing the full stationary distribution
on the $(x,z)$-grid. As for the simulated maximum likelihood estimator,
smoothness is barred by the interpolation required, and we therefore
use the same Nelder-Mead type optimizer for maximizing $\bar{l}(\theta)$.
Moreover, we rely on parametric bootstrap with (\ref{eq:stoch_model1}-\ref{eq:stoch_model3})
as data generating process for approximate standard errors. 

Modulo Monte Carlo approximation error, it is seen that $\hat{\theta}$
maximizes $\sum_{t=2}^{T}\log\pi(p_{t}|p_{1},\dots,p_{t-1})$, whereas
$\bar{\theta}$ maximizes $\sum_{t=2}^{T}\log\pi(p_{t}|p_{t-1})$.
A number of factors determine how much the two differ. In particular
are $p_{t}|p_{t-1}$and $p_{t}|p_{1},\dots,p_{t-1}$ equal in distribution
when the non-negativity constraint is binding, and therefore one would
primarily expect the two estimators to differ when $\delta$ is small
and stock-outs are infrequent. Moreover, one would expect that the
two would differ more at higher sampling frequencies or with values
of $\rho$ closer to 1 since then $p_{1},\dots,p_{t-2}$ are more
informative with respect to $p_{t}$. 

\section{Simulation study}

To study the performance of the proposed particle filter-based simulated
ML estimator (SML) based on $\hat{l}(\theta)$, and also to compare
with the composite ML (CML) estimator based on $\bar{l}(\theta)$,
we conduct a simulation study. It should be noted that data are simulated
using the numerical solution $\hat{f}$, and biases incurred by using
$\hat{f}$ instead of $f$ in the estimation are consequently not
visible in this study. See section \ref{subsec:Diagnostics} for a
study of the sensitivity of parameter estimates to the grid size.
Throughout, we use $N=4096$ particles in the filter, and the real
interest rate $r$ is chosen to correspond to a 5\% yearly rate. Each
experiment is repeated 100 times and the true parameters in the data
generating process (\ref{eq:stoch_model1}-\ref{eq:stoch_model3})
are taken to resemble those found in empirical applications. Throughout,
the optimizations performed under SML and CML are started at the true
parameters of the data generating process.

\subsection{Monthly data experiment\label{subsec:Monthly-data-experiment}}

\begin{table}
\begin{centering}
\begin{tabular}{llccccc}
\hline 
 &  & $\rho$ & $a$ & $b$ & $\delta$ & (q-)log-like\tabularnewline
\hline 
True parameters &  & 0.97 & 1.5 & -0.4  & 0.02 & \tabularnewline
\hline 
\multicolumn{7}{c}{$T=250,\;\;$$\tau_{SML}=2.4$s, $\tau_{CML}=7.1$s}\tabularnewline
\hline 
SML & Bias & -0.0146 & -0.4821 & -0.2173 &  0.0011 & \tabularnewline
 & Std.dev. & 0.0232 & 1.5897 & 0.9636 &  0.0078 & \tabularnewline
 & RMSE & 0.0273 & 1.6536 & 0.9831 &  0.0078 & \tabularnewline
 & MC.Std.dev & 0.0010 & 0.0315 & 0.0069 & 0.0007 &  0.0376\tabularnewline
 &  &  &  &  &  & \tabularnewline
CML & Bias & -0.0264 & -0.4200 & -0.3923 & 0.0054 & \tabularnewline
 & Std.dev. & 0.0397 & 3.3513 & 1.3833 & 0.0086 & \tabularnewline
 & RMSE & 0.0475 & 3.3606 & 1.4311  & 0.0102 & \tabularnewline
 & MC.Std.dev & 0.0025 & 0.0841 & 0.0269 & 0.0005 &  0.0982\tabularnewline
\hline 
\multicolumn{7}{c}{$T=500,\;\;$$\tau_{SML}=3.0$s, $\tau_{CML}=11.8$s}\tabularnewline
\hline 
SML & Bias & -0.0050 & -0.1110 & -0.0340 &  0.0005 & \tabularnewline
 & Std.dev. & 0.0124 & 0.5020 & 0.2169 &  0.0050 & \tabularnewline
 & RMSE & 0.0133 & 0.5116 & 0.2185 &  0.0050 & \tabularnewline
 & MC.Std.dev & 0.0004 & 0.0095 & 0.0059 & 0.0002 &  0.0522\tabularnewline
 &  &  &  &  &  & \tabularnewline
CML & Bias & -0.0185 & -0.1922 & -0.3953 & 0.0048 & \tabularnewline
 & Std.dev. & 0.0256 & 1.6313 & 0.8874 & 0.0071 & \tabularnewline
 & RMSE & 0.0315 & 1.6344 & 0.9673 & 0.0085 & \tabularnewline
 & MC.Std.dev & 0.0011 & 0.0195 & 0.0060 & 0.0005 &  0.0884\tabularnewline
\hline 
\multicolumn{7}{c}{$T=1000,\;\;$$\tau_{SML}=3.9$s, $\tau_{CML}=19.5$s}\tabularnewline
\hline 
SML & Bias & -0.0021 & -0.0258 & 0.0028 &  0.0004 & \tabularnewline
 & Std.dev. & 0.0065 & 0.2280 & 0.0664 &  0.0031 & \tabularnewline
 & RMSE & 0.0068 & 0.2284 & 0.0661 &  0.0031 & \tabularnewline
 & MC.Std.dev & 0.0003 & 0.0084 & 0.0026 & 0.0002 &  0.0191\tabularnewline
 &  &  &  &  &  & \tabularnewline
CML & Bias & -0.0109 & -0.1408 & -0.3801 & 0.0047 & \tabularnewline
 & Std.dev. & 0.0101 & 1.4722 & 0.7430 & 0.0054 & \tabularnewline
 & RMSE & 0.0149 & 1.4715 & 0.8313 & 0.0071 & \tabularnewline
 & MC.Std.dev & 0.0020 & 0.1067 & 0.0280 & 0.0007 &  0.0970\tabularnewline
\hline 
\end{tabular}
\par\end{centering}
{\footnotesize{}Note: True parameters indicated in the uppermost row
and $r$ corresponding to monthly data. Bias, statistical standard
errors (Std.dev) and root mean squared errors (RMSE) are based on
100 replicas, with all replicas for SML converged. For CML, the number
of failed and ignored replica were 1 ($T=250)$, 1 ($T=500$) and
2 ($T=1000$). For both SML and CML, different random number seeds
were used in all replica, and $N=4096$ particles were employed for
SML. Monte Carlo Standard errors (MC.Std.dev) are calculated from
10 repeated estimations to a simulated data set with different random
number seeds. $\tau_{SML}$ and $\tau_{CML}$ denote the mean (clock-)time
of evaluating a (quasi-)log-likelihood function on a Dell Latitude
E6510 laptop with an Intel Core i7 CPU 1.73 GHz CPU with 8 cores running
Linux.}{\footnotesize \par}
\centering{}\caption{\label{tab:Simulation-study-month}Simulation study for the model
(\ref{eq:stoch_model1}-\ref{eq:stoch_model3}), monthly data experiment. }
\end{table}
In the first simulation study, we consider prices at an equivalent
monthly frequency and true parameters inspired by natural gas real
data. The bias, standard deviation and root mean square error (RMSE)
are reported in Table \ref{tab:Simulation-study-month} for different
sample sizes $T=\left\{ 250,500,1000\right\} $. For the SML estimator,
some biases are seen for the shorter sample size $T=250$, but these
biases seem to diminish for the larger sample sizes at a rate that
is largely consistent with- or faster than what standard maximum likelihood
theory would predict. Largely the same effects are seen for the standard
deviations and RMSEs. Looking at the magnitudes of the RMSEs, we see
that the parameters $\rho$ and $\delta$, which in large part determine
the temporal dependence structure, are estimated with relatively good
precision. On the other hand, $a$ and \textbf{$b$}, which as a rule
of thumb correspond to marginal location and scale of the price process
sees relatively large RMSEs, in particular for the smaller sample
sizes. This is related to the fact that the estimation of these parameters
is strongly influenced by the price spikes, and the number and severity
of price spikes vary substantially over the simulated data sets. 

For CML, we obtain RMSEs that are consistently larger than those of
SML, with differences being largest for $a$ and $b$ at the larger
sample sizes. Moreover, in line with \citet{Michaelides2000}, substantial
biases that vanishes only very slowly as $T$ increases are seen for
CML. Looking at relative computing times and errors induced by using
Monte Carlo methods, we see that SML is substantially less expensive
to evaluate and also produces consistently smaller Monte Carlo errors
than those of CML. 

\subsection{Weekly data experiment}

Our second simulation study involves prices at an equivalent weekly
frequency and true parameters similar to those obtained for export
prices from Norway on fresh farmed Atlantic Salmon\footnote{The real data used are between week 1, 1995 and week 39, 2012. The
prices were normalized to mean 1. Prices can be found at http://www.nosclearing.com/}. The design of the experiment is otherwise equal to the monthly data
experiment and the results are provided in Table \ref{tab:Simulation-study-week}.
As in the previous experiment, SML appear to work satisfactory, with
small biases for the larger sample sizes and RMSEs decaying as $T^{-1/2}$
or faster. In line with what was found for the monthly data experiment,
it is seen that the temporal dependence parameters $\rho$ and $\delta$
are estimated with good precision in all cases, whereas the location
and scale parameters $a$ and $b$ require more data to be accurately
determined. 

CML has again mostly larger RMSEs than SML in all cases except for
$b$ in the $T=250$ case, for which SML results are dominated by
a single extreme replica. In particular we see a non-vanishing bias
for $a$-parameter for CML that again mirrors what was found by \citet{Michaelides2000}.
The differences in computing time and Monte Carlo standard errors
are also consistent with the previous experiment, with both higher
computational cost and higher Monte Carlo standard errors associated
with CML. We therefore conclude that also in this setup, SML is the
better overall estimator. 
\begin{table}
\begin{centering}
\begin{tabular}{llccccc}
\hline 
 &  & $\rho$ & $a$ & $b$ & $\delta$ & (q-)log-like\tabularnewline
\hline 
True Parameters &  & 0.99 & 1.65 & -0.09 & 0.0035 & \tabularnewline
\hline 
\multicolumn{7}{c}{$T=250,\;\;$$\tau_{SML}=2.5$s, $\tau_{CML}=7.3$s}\tabularnewline
\hline 
SML & Bias & -0.0079 & -0.1401 & -0.1473 &  0.0004 & \tabularnewline
 & Std.dev. & 0.0294 & 0.3970 & 1.2646 &  0.0025 & \tabularnewline
 & RMSE & 0.0303 & 0.4191 & 1.2669 &  0.0026 & \tabularnewline
 & MC.Std.dev & 0.0002 & 0.0044 & 0.0008 & 2.8e-5 & 0.0096\tabularnewline
 &  &  &  &  &  & \tabularnewline
CML & Bias & -0.0179 & 0.1668 & -0.0966 & 0.0024 & \tabularnewline
 & Std.dev. & 0.0268 & 0.6931 & 0.1865 & 0.0033 & \tabularnewline
 & RMSE & 0.0321 & 0.7095 & 0.2092 & 0.0041 & \tabularnewline
 & MC.Std.dev & 0.0011 & 0.0428 & 0.0028 & 0.0004 & 0.1237\tabularnewline
\hline 
\multicolumn{7}{c}{$T=500,\;\;$$\tau_{SML}=2.9$s, $\tau_{CML}=12.1$s}\tabularnewline
\hline 
SML & Bias & -0.0021 & -0.0024 & -0.0107  &  0.0003 & \tabularnewline
 & Std.dev. & 0.0068 & 0.2189 & 0.0470 &  0.0013 & \tabularnewline
 & RMSE & 0.0071 & 0.2178 & 0.0480 &  0.0013 & \tabularnewline
 & MC.Std.dev & 0.0001 & 0.0010 & 0.0005 & 3.1e-5 & 0.0120\tabularnewline
 &  &  &  &  &  & \tabularnewline
CML & Bias & -0.0174 & 0.2663 & -0.1654 & 0.0025 & \tabularnewline
 & Std.dev. & 0.0562 & 0.5712 & 0.6448 & 0.0030 & \tabularnewline
 & RMSE & 0.0585 & 0.6277 & 0.6626 & 0.0039 & \tabularnewline
 & MC.Std.dev & 0.0046 & 0.2099 & 0.0168 & 0.0008 & 0.9993\tabularnewline
\hline 
\multicolumn{7}{c}{$T=1000,\;\;$$\tau_{SML}=3.9$s, $\tau_{CML}=19.9$s}\tabularnewline
\hline 
SML & Bias & -0.0006 & -0.0083 & -0.0015 &  -8.0e-7 & \tabularnewline
 & Std.dev. & 0.0023 & 0.1590 & 0.0123 &  0.0007 & \tabularnewline
 & RMSE & 0.0024 & 0.1584 & 0.0123 &  0.0007 & \tabularnewline
 & MC.Std.dev & 1.5e-5 & 0.0057 & 0.0002 & 1.3e-5 & 0.0325\tabularnewline
 &  &  &  &  &  & \tabularnewline
CML & Bias & -0.0060 & 0.4690 & -0.0754 & 0.0025 & \tabularnewline
 & Std.dev. & 0.0063 & 0.5164 & 0.1470 & 0.0023 & \tabularnewline
 & RMSE & 0.0087 & 0.6956 & 0.1646 & 0.0034 & \tabularnewline
 & MC.Std.dev & 0.0063 & 0.3497 & 0.0434 & 0.0004 & 0.5395\tabularnewline
\hline 
\end{tabular}
\par\end{centering}
{\footnotesize{}Note: True parameters indicated in the uppermost row
and $r$ corresponding to weekly data. Bias, statistical standard
errors (Std.dev) and root mean squared errors (RMSE) are based on
100 replicas, with all replicas for SML converged. For both SML and
CML, different random number seeds were used in all replica, and $N=4096$
particles were employed for SML. Monte Carlo Standard errors (MC.Std.dev)
are calculated from 10 repeated estimations to a simulated data set
with different random number seeds. $\tau_{SML}$ and $\tau_{CML}$
denote the mean (clock-)time of evaluating a (quasi-)log-likelihood
function on a Dell Latitude E6510 laptop with an Intel Core i7 CPU
1.73 GHz CPU with 8 cores running Linux.}{\footnotesize \par}
\centering{}\caption{\label{tab:Simulation-study-week}Simulation study for the model (\ref{eq:stoch_model1}-\ref{eq:stoch_model3}),
weekly data experiment. }
\end{table}

\subsection{Yearly data experiment\label{subsec:Yearly-data-experiment}}

Earlier applications of the storage model \citep{Deaton1995,Deaton1996}
have relied on a yearly period modeling and yearly data. This leads
to smaller supply shock autocorrelation, and therefore provide a situation
that is somewhat less in favor of SML. In the yearly data experiment
we consider sample sizes of $T=100,\text{\;}200$ years of data and
true dynamics equal to the parameters obtained by \citet{Deaton1996}
for Tin using yearly data between 1900 and 1987. Otherwise the design
of the experiment is equal to the monthly data experiment and results
are presented in Table \ref{tab:Simulation-study-year}.
\begin{table}
\begin{centering}
\begin{tabular}{llccccc}
\hline 
 &  & $\rho$ & $a$ & $b$ & $\delta$ & (q-)log-like\tabularnewline
\hline 
True Parameters &  & 0.918 & 0.223 & -0.038 & 0.046 & \tabularnewline
\hline 
\multicolumn{7}{c}{$T=100,\;\;$$\tau_{SML}=2.1$s, $\tau_{CML}=4.0$s}\tabularnewline
\hline 
SML & Bias &   -0.0047 & 0.0007 & -0.0008 & 0.0054 &      \tabularnewline
 & Std.dev. &     0.0338 &     0.0391 &     0.0063 &     0.0234 &          \tabularnewline
 & RMSE &     0.0339 &     0.0389 &     0.0063 &     0.0238 &          \tabularnewline
 & MC.Std.dev &     0.0003 &     0.0002 &   3.4e-5 &     0.0001 &     0.0117\tabularnewline
 &  &  &  &  &  & \tabularnewline
CML & Bias &    -0.0197 &     0.0258 &    -0.0200 &    -0.0040 &          \tabularnewline
 & Std.dev. &     0.0338 &     0.0445 &     0.0122 &     0.0318 &          \tabularnewline
 & RMSE &     0.0390 &     0.0512 &     0.0234 &     0.0319 &          \tabularnewline
 & MC.Std.dev &     0.0021 &     0.0015 &     0.0012 &     0.0008 &     0.0469\tabularnewline
\hline 
\multicolumn{7}{c}{$T=200,\tau_{SML}=2.3$s, $\tau_{CML}=5.3$s}\tabularnewline
\hline 
SML & Bias & -0.0026 & 0.0008 &  -0.0011 & -0.0020 &       \tabularnewline
 & Std.dev. &     0.0219 &    0.0293 &     0.0044 &     0.0144 &          \tabularnewline
 & RMSE &     0.0220 &     0.0292 &     0.0045 &     0.0145 &          \tabularnewline
 & MC.Std.dev &     0.0002 &     0.0010 &     3.2e-5 &     0.0001 &     0.0447\tabularnewline
 &  &  &  &  &  & \tabularnewline
CML & Bias &    -0.0157 &     0.0252 &    -0.0181 &    -0.0091 &          \tabularnewline
 & Std.dev. &     0.0257 &     0.0359 &     0.0071 &     0.0140 &          \tabularnewline
 & RMSE &     0.0300 &     0.0437 &     0.0195 &     0.0166 &          \tabularnewline
 & MC.Std.dev &     0.0006 &     0.0016 &     0.0002 &     0.0009 &     0.1350\tabularnewline
\hline 
\end{tabular}
\par\end{centering}
{\footnotesize{}Note: True parameters indicated in the uppermost row
and $r$ corresponding to yearly data. Bias, statistical standard
errors (Std.dev) and root mean squared errors (RMSE) are based on
100 replicas, with all replicas for SML converged. For both SML and
CML, different random number seeds were used in all replica, and $N=4096$
particles were employed for SML. Monte Carlo Standard errors (MC.Std.dev)
are calculated from 10 repeated estimations to a simulated data set
with different random number seeds. $\tau_{SML}$ and $\tau_{CML}$
denote the mean time of evaluating a (quasi-)log-likelihood function
on a 2016 imac with an 3.1 Ghz Intel Core i5. Only a single tread
was used.}{\footnotesize \par}
\centering{}\caption{\label{tab:Simulation-study-year}Simulation study for the model (\ref{eq:stoch_model1}-\ref{eq:stoch_model3}),
yearly data experiment. }
\end{table}

Also here it is seen that SML works very satisfactory with small biases
throughout. Moreover, the Monte Carlo standard errors are substantially
smaller than the statistical standard errors. In this situation, it
is seen that the biases for CML vanishes faster than the monthly and
weekly situations. Still for CML, we see larger biases and RMSEs than
for SML. Taking into account that CML is more computationally costly
than SML, we conclude that SML is the preferred estimator, even for
data at a yearly frequency.

\subsection{Effects of estimation bias on implied price characteristics}

Table \ref{tab:biastable} 
\begin{table}
\begin{centering}
\begin{tabular}{ccccccc}
\hline 
 & $E\left(p_{t}\right)$ & $SD\left(p_{t}\right)$ & $SK\left(p_{t}\right)$ & $KU\left(p_{t}\right)$ & $AC_{1}\left(p_{t}\right)$ & P(stock out)\tabularnewline
\hline 
 &  &  &  &  &  & \tabularnewline
\hline 
\multicolumn{7}{c}{Monthly data experiment}\tabularnewline
\hline 
 &  &  &  &  &  & \tabularnewline
True dynamics & 0.8583 & 0.6752 & 2.3978 & 10.6107 & 0.9677 & 0.0423\tabularnewline
 &  &  &  &  &  & \tabularnewline
T=250, SML & 0.6641 & 0.5045 & 3.2359 & 19.2188 & 0.9448 & 0.0298\tabularnewline
T=500, SML & 0.8047 & 0.6056 & 2.5399 & 11.8934 & 0.9608 & 0.0412\tabularnewline
T=1000, SML & 0.8517 & 0.6473 & 2.3629 & 10.4207 & 0.9653 & 0.0454\tabularnewline
T=250, CML & 0.7440 & 0.5582 & 3.2888 & 20.0495 & 0.9294 & 0.0359\tabularnewline
T=500, CML & 0.8396 & 0.6599 & 3.2027 & 18.5543 & 0.9397 & 0.0377\tabularnewline
T=1000, CML & 0.8423 & 0.7309 & 3.2759 & 18.7207 & 0.9492 & 0.0333\tabularnewline
 &  &  &  &  &  & \tabularnewline
\hline 
\multicolumn{7}{c}{Weekly data experiment}\tabularnewline
\hline 
 &  &  &  &  &  & \tabularnewline
True dynamics & 1.2018 & 0.4022 & 1.0890 & 4.3193 & 0.9909 & 0.0119\tabularnewline
 &  &  &  &  &  & \tabularnewline
T=250, SML & 1.0349 & 0.3683 & 1.8842 & 9.3327 & 0.9817 & 0.0132\tabularnewline
T=500, SML & 1.1989 & 0.3820 & 1.1783 & 4.6949 & 0.9887 & 0.0145\tabularnewline
T=1000, SML & 1.1992 & 0.3888 & 1.1095 & 4.4118 & 0.9902 & 0.0111\tabularnewline
T=250, CML & 1.3514 & 0.3570 & 1.4318 & 6.1169 & 0.9714 & 0.0261\tabularnewline
T=500, CML & 1.3727 & 0.4125 & 1.6494 & 7.4317 & 0.9716 & 0.0257\tabularnewline
T=1000, CML & 1.5070 & 0.5417 & 1.2826 & 4.9978 & 0.9848 & 0.0244\tabularnewline
 &  &  &  &  &  & \tabularnewline
\hline 
\multicolumn{7}{c}{Yearly data experiment}\tabularnewline
\hline 
 &  &  &  &  &  & \tabularnewline
True dynamics & 0.1922 & 0.0875 & 0.4582 & 2.8062 & 0.9063 & 0.0733\tabularnewline
 &  &  &  &  &  & \tabularnewline
T=100, SML &     0.1941 &     0.0871 &     0.4462 &     2.8027 &     0.8991 &     0.0846\tabularnewline
T=200, SML &     0.1913 &     0.0876 &     0.4826 &     2.8363 &     0.9050 &     0.0771\tabularnewline
T=100, CML &     0.1928 &     0.1058 &     0.8067 &     3.3511 &     0.8949 &     0.0688\tabularnewline
T=200, CML &     0.1921 &     0.1041 &     0.8074 &     3.3503 &     0.8997 &     0.0717\tabularnewline
 &  &  &  &  &  & \tabularnewline
\hline 
\end{tabular}
\par\end{centering}
{\footnotesize{}Note: All figures are calculated from one million
periods of simulated dynamics. \textquotedblleft True dynamics\textquotedblright{}
are calculated from dynamics with parameters equal to those of the
\textquotedblleft True parameters\textquotedblright{} rows of Tables
1, 2 and 3. The remaining figures are calculated from dynamics with
parameters equal to \textquotedblleft True parameters\textquotedblright{}
plus \textquotedblleft Bias\textquotedblright{} taken from Tables
1, 2 and 3. $SD$ denotes standard deviations, $SK$ skewness, $KU$
kurtosis, $AC_{1}$ first order autocorrelation and P(stock out) denotes
the marginal probability of being in a stock-out state, i.e. $P\left(x_{t}\right)=f\left(x_{t},z_{t}\right)$.}{\footnotesize \par}

\caption{\label{tab:biastable}The effect of estimation bias on price characteristics }
\end{table}
 investigates the effects of the estimation bias on implied price
characteristics for the monthly, weekly and yearly data experiments.
Using the true and ``biased'' parameters from the estimations in
table \ref{tab:Simulation-study-month} and \ref{tab:Simulation-study-week},
we calculate and compare the mean, standard deviation, skewness, kurtosis
and first order autocorrelation of model simulated prices. We also
calculate the marginal probability a stock out as implied by the estimated
models. This is done using both the monthly and weekly data experiment
parameters. What is immediately clear from the table is that for the
SML estimator, bias declines over all statistics as more price observations
become available. This is not so for the CML estimator. For some of
the statistics, for instance the mean in the weekly data experiment
or the kurtosis in the monthly data experiment, the bias increases
as more data becomes available. Overall the SML estimator displays
favorable large sample properties and improved precision over the
CML estimator. We also note that the CML estimator consistently underestimates
the price autocorrelation. This is relevant as low implied price autocorrelation
has been pointed to as a weakness of the competitive storage model
\citep{Deaton1996,cafiero2011empirical}.

\subsection{Robustness to misspecification}

A reason for opting for composite likelihood methods is that in many
cases such methods may be more robust to model misspecification than
methods based on the full likelihood. See e.g. \citet{doi:10.1093/biomet/91.3.729}
or \citet{10.2307/24309261}, section 4.3 for a detailed discussion
of the robustness properties associated with composite likelihood.
To compare CML and SML under misspecification, we carry out a Monte
Carlo study where in the data generating process, $\eta_{t}$ in (\ref{eq:stoch_model1})
is taken to be a scaled $t_{4}$-distribution with unit variance.
The estimation methodology is left unchanged, and we consider the
an identical setup as for the yearly data experiment discussed in
Section \ref{subsec:Yearly-data-experiment}. 
\begin{table}
\begin{centering}
\begin{tabular}{llcccc}
\hline 
 &  & $\rho$ & $a$ & $b$ & $\delta$\tabularnewline
\hline 
True parameters &  & 0.918 & 0.223 & -0.038  & 0.046\tabularnewline
\hline 
\multicolumn{6}{c}{$T=100,\;\;$$\tau_{SML}=2.1$s, $\tau_{CML}=4.0$s}\tabularnewline
\hline 
SML & Bias & -0.0095 & 0.0041 & -0.0017 & -0.0029\tabularnewline
 & Std.dev. &     0.0375 &     0.0355 &     0.0085 &     0.0206\tabularnewline
 & RMSE &     0.0385 &     0.0356 &     0.0086 &     0.0207\tabularnewline
 &  &  &  &  & \tabularnewline
CML & Bias &    -0.0362 &     0.0237 &    -0.0220 &    -0.0072\tabularnewline
 & Std.dev. &     0.0442 &     0.0427 &     0.0186 &     0.0284\tabularnewline
 & RMSE &     0.0570 &     0.0487 &     0.0288 &     0.0292\tabularnewline
\hline 
\multicolumn{6}{c}{$T=200,\;\;$$\tau_{SML}=2.3$s, $\tau_{CML}=5.3$s}\tabularnewline
\hline 
SML & Bias & 0.0005 & 0.0015 & 0.0005 & -0.0049\tabularnewline
 & Std.dev. &     0.0207 &     0.0265 &     0.0048 &     0.0142\tabularnewline
 & RMSE &     0.0206 &     0.0265 &     0.0048 &     0.0149\tabularnewline
 &  &  &  &  & \tabularnewline
CML & Bias &    -0.0218 &     0.0218 &    -0.0170 &    -0.0034\tabularnewline
 & Std.dev. &     0.0333 &     0.0328 &     0.0127 &     0.0201\tabularnewline
 & RMSE &     0.0396 &     0.0392 &     0.0212 &     0.0203\tabularnewline
\hline 
\end{tabular}
\par\end{centering}
\begin{raggedright}
{\footnotesize{}Note: True parameters indicated in the uppermost row
and $r$ corresponding to yearly data. Bias, statistical standard
errors (Std.dev) and root mean squared errors (RMSE) are based on
100 replicas, with all replicas for SML converged. For both SML and
CML, different random number seeds were used in all replica, and $N=4096$
particles were employed for SML. $\tau_{SML}$ and $\tau_{CML}$ denote
the mean time of evaluating a (quasi-)log-likelihood function on a
2016 imac with an 3.1 Ghz Intel Core i5. Only a single tread was used.}
\par\end{raggedright}{\footnotesize \par}
\caption{\label{tab:Simulation-study-for-misspec}Simulation study for the
model (\ref{eq:stoch_model1}-\ref{eq:stoch_model3}) subject to misspecification
in $\eta_{t}$.}

\end{table}

The results are presented in Table \ref{tab:Simulation-study-for-misspec}.
It is seen that the changes in performance under misspecification
is relatively minor for both methods (i.e. relative to Table \ref{tab:Simulation-study-year}).
In particular, we see that the relative performance between SML and
CML is largely unchanged when misspecified $\eta_{t}$ is added to
the data generating process. Thus, at least for this situation and
a heavy-tailed form of misspecification, the benefits from employing
the potentially more robust CML are too small to outweigh the statistical
accuracy stemming from using a full likelihood specification as in
SML.

\section{Empirical Application}

To illustrate our estimation procedure with real data we apply it
to monthly frequency observations on natural gas spot prices. To compare
model fit and precision of parameter estimates we apply both our filter-based
estimator and the composite likelihood estimator. One way to empirically
asses the relevance of speculative storage model is to compare the
storage model fit to a benchmark linear AR(1) model. It is straightforward
to show that zero storage (which might occur if the cost of storage
is sufficiently large) in the storage model implies that prices evolve
as
\begin{equation}
p_{t+1}=a+\rho(p_{t}-a)+b\epsilon_{t+1},\;\epsilon_{t}\sim\text{i.i.d. }N(0,1).\label{eq:benchmark_model}
\end{equation}
Rejecting the linear AR(1) model in favor of the storage model does
not imply that the storage model is the ``true'' model, but provides
support for characteristics consistent with speculative storage over
a model where prices are explained by a linear first order process.
One reason why a linear AR(1) model might be rejected in favor of
the storage model is stochastic volatility in the data. To investigate
stochastic volatility we also estimate the linear AR(1) model with
GARCH(1,1) errors. Finally, given the storage model predicts a two-state
regime switching type price dynamics, we estimate a Markov switching
(MS) model, where the AR(1) model in equation \ref{eq:benchmark_model}
is allowed to change between two regimes as determined by a latent
two-state Markov process with constant transition probabilities. 

Comparing the fit of the competitive storage model to reduced form
time-series models is a strong test. The reduced form models (such
as the linear AR(1) model or the GARCH models) are specifically designed
to account for specific features of the data (such as first-order
autocorrelation or ARCH effects). The value of such a comparison is
that we can identify price features that the model is able to account
for, and where improvements are needed. Finally, we note that all
models considered are estimated using commodity price data only.

\subsection{Application to Natural Gas at Henry Hub}

The natural gas price in this analysis is the spot price at the Henry
Hub terminal in Louisiana. Prices are denoted in US\$ per thousand
cubic meters of gas, and covers the period 1991 M1 to 2012 M6. Prices
can be obtained from http://www.imf.org/external/np/res/commod/index.aspx.
Natural gas prices are influenced by a range of factors such as the
price of oil, weather, seasonality in demand, shut-in production and
storage \citep{brown2008drives}. In regards to storage, inventories
play an important role in smoothing production and balancing demand\textendash supply
conditions. The release of information on inventory levels is known
to generate considerable volatility in prices \citep{mu2007weather}.
\citet{chiou2006commodity} further demonstrate that the convenience
yield (measured as the difference between the spot and forward natural
gas price) is negatively related to natural gas inventory levels.
This suggests that abnormally high prices are related to low inventories,
as implied by the storage model. 

The storage model is arguably too simple to capture the full complexities
of the real market. Effects related to energy substitution (the oil
price effect) and seasonality in demand is not specifically modeled,
and is likely captured by the supply shocks. However, if speculative
storage effects are present to any substantial degree, inference should
favor the storage model over the linear AR(1) model (equation \ref{eq:benchmark_model}).
An added benefit of analyzing the natural gas market is that detailed
information on natural gas storage is available. This means we can
compare the model implied filtered storage, derived using only the
price data, to actual storage levels. A strong correspondence between
these series will provide support for the relevance of the storage
model and the estimation procedure. 

Prior to estimation the price data was normalized to unit mean. Parameter
estimates are provided in Table \ref{tab:Parameter-estimates-natgas}.
\begin{table}
\begin{centering}
\begin{tabular}{lcccccc}
\hline 
 & $\rho$ & $a$ & $b$ & $\delta$ &  & log-likelihood\tabularnewline
\hline 
\multicolumn{7}{c}{SML}\tabularnewline
\hline 
 &  &  &  &  &  & \tabularnewline
Estimate & 0.968 & 1.471 & -0.408 & 0.0212 &  &  194.32\tabularnewline
MC Std.Dev & 2.9e-5 & 4.2e-4 & 1.8e-4 & 1.0e-5 &  & 4.6e-3\tabularnewline
Statistical S.E. & 0.0265 & 1.457 & 0.489 & 0.0083 &  & \tabularnewline
 &  &  &  &  &  & \tabularnewline
\hline 
\multicolumn{7}{c}{CML}\tabularnewline
\hline 
 &  &  &  &  &  & \tabularnewline
Estimate & 0.963 & 2.075 & -0.599 & 0.0275 &  & 192.19\tabularnewline
MC Std.Dev & 0.0039 & 0.237 & 0.0759 & 0.0016 &  & 0.288\tabularnewline
Statistical S.E. & 0.0465 & 1.213 & 1.360 & 0.0102 &  & \tabularnewline
 &  &  &  &  &  & \tabularnewline
\hline 
\multicolumn{7}{c}{Benchmark AR(1) model (\ref{eq:benchmark_model})}\tabularnewline
\hline 
 &  &  &  &  &  & \tabularnewline
Estimate & 0.950 & 1.021 & -0.188 &  &  & 65.34\tabularnewline
Statistical S.E. & 0.0192 & 0.625 & 0.0083 &  &  & \tabularnewline
 &  &  &  &  &  & \tabularnewline
\hline 
\multicolumn{7}{c}{Markov-Switching AR(1) model, }\tabularnewline
\multicolumn{7}{c}{$P(\text{Regime 1 next period}|\text{Regime 1 current period)=}0.951$,}\tabularnewline
\multicolumn{7}{c}{$P(\text{Regime 1 next period}|\text{Regime 2 current period)=}0.081$}\tabularnewline
\hline 
 &  &  &  &  &  & \tabularnewline
Estimate (Regime 1) & 0.887 & 0.549 & -0.064 &  &  & 164.09\tabularnewline
Statistical S.E. (Regime 1) & 0.0164 & 0.014 & 0.0048 &  &  & \tabularnewline
Estimate (Regime 2) & 0.861 & 1.725 & -0.28 &  &  & \tabularnewline
Statistical S.E. (Regime 2) & 0.0492 & 0.884 & 0.022 &  &  & \tabularnewline
 &  &  &  &  &  & \tabularnewline
\hline 
\multicolumn{7}{c}{AR(1)-GARCH(1,1) (cond. variance model: $\sigma_{t}^{2}=\alpha_{0}+\alpha_{1}\epsilon_{t-1}^{2}+\beta_{1}\sigma_{t-1}^{2})$ }\tabularnewline
 & $\rho$ & $a$ & $\alpha_{0}$ & $\alpha_{1}$ & $\beta_{1}$ & \tabularnewline
\hline 
 &  &  &  &  &  & \tabularnewline
Estimate & 0.967  & 0.520  & 4.1e-4 & 0.046 & 0.667 & 148.79\tabularnewline
Statistical S.E. & 0.0155 & 0.396 & 2.5e-4 & 0.095 & 0.044 & \tabularnewline
 &  &  &  &  &  & \tabularnewline
\hline 
\end{tabular}
\par\end{centering}
{\footnotesize{}Note: Standard errors are based on 100 parametric
bootstrap replicas, and MC standard errors are based on fitting the
model to the real data 30 times with different random number seeds
in the particle filter. The benchmark AR(1) model was fitted using
maximum likelihood, and statistical standard errors are based on observed
Fisher information.}{\footnotesize \par}
\centering{}\caption{\label{tab:Parameter-estimates-natgas}Parameter estimates for the
natural gas data. }
\end{table}
 We see that SML and CML produce similar results, but that SML has
smaller Monte Carlo standard errors. The statistical- and Monte Carlo
standard error for the real data are largely consistent with those
found in the simulation study reported in Table \ref{tab:Simulation-study-month},
$T=250$. Compared to the AR(1) benchmark, the storage model performs
substantially better in terms of model fit. This implies that not
all of the model fit comes from the exogenous latent AR(1) shock;
the economic model appears relevant in terms of explaining observed
characteristics. 

Adding GARCH(1,1) errors to the AR(1) model gives an improved fit
with a log-likelihood of 148.79, where the AR(1)-GARCH(1,1) model
has one more parameter than the storage model. However, the log-likelihood
is still below the fit of the storage model. Due to the non-nested
nature of the AR(1)-GARCH(1,1) model and the storage model (\ref{eq:stoch_model1}-\ref{eq:stoch_model3}),
we also computed the log-likelihood ratios of the storage model against
the AR(1)-GARCH(1,1) model for the 100 parametric bootstrap replica
reported in Table \ref{tab:Parameter-estimates-natgas} \footnote{I.e. (\ref{eq:stoch_model1}-\ref{eq:stoch_model3}) with parameters
found for SML is the data-generating process.}. We find that the observed log-likelihood ratio ($2(194.32-148.79)=91.06$)
fall between the 61st and 62nd simulated likelihood ratio when the
storage model is the true model, which indicate that the observed
log-likelihood ratio is consistent with the storage model being the
``true'' model. 

The MS-AR(1) produces a fit with log-likelihood of 164.09, which is
again substantially below the structural model while containing four
more parameters. To assess this number, we fitted the MS-AR(1) model
to the simulated data under the storage model as described above.
We find that the observed likelihood ratio ($2(194.32-163.02)=60.46$)
fall between the 13th and 14th simulated likelihood ratio when the
storage model is the true model, which is again consistent with the
storage model being the better model. 

Finally, we fitted model (\ref{eq:stoch_model1}-\ref{eq:stoch_model3})
with iid supply shock (i.e. $\rho=0$) to compare with a methodology
similar to that of \citet{cafiero2011empirical}. We used the same
price function solver and conditionally Gaussian transition densities
for the price process so that the iid supply shock model is nested
under the general model (\ref{eq:stoch_model1}-\ref{eq:stoch_model3}).
We obtain a log-likelihood of 190.79 for the iid supply shock model,
which correspond to a rejection of the iid model against the general
storage model (\ref{eq:stoch_model1}-\ref{eq:stoch_model3}) with
$p$-value 0.008. It is also worth noticing that the fitted price
function has a very large negative $b=-189$, and therefore we regard
this model fit as nonsensical. 

\begin{figure}
\begin{centering}
\includegraphics[scale=0.5]{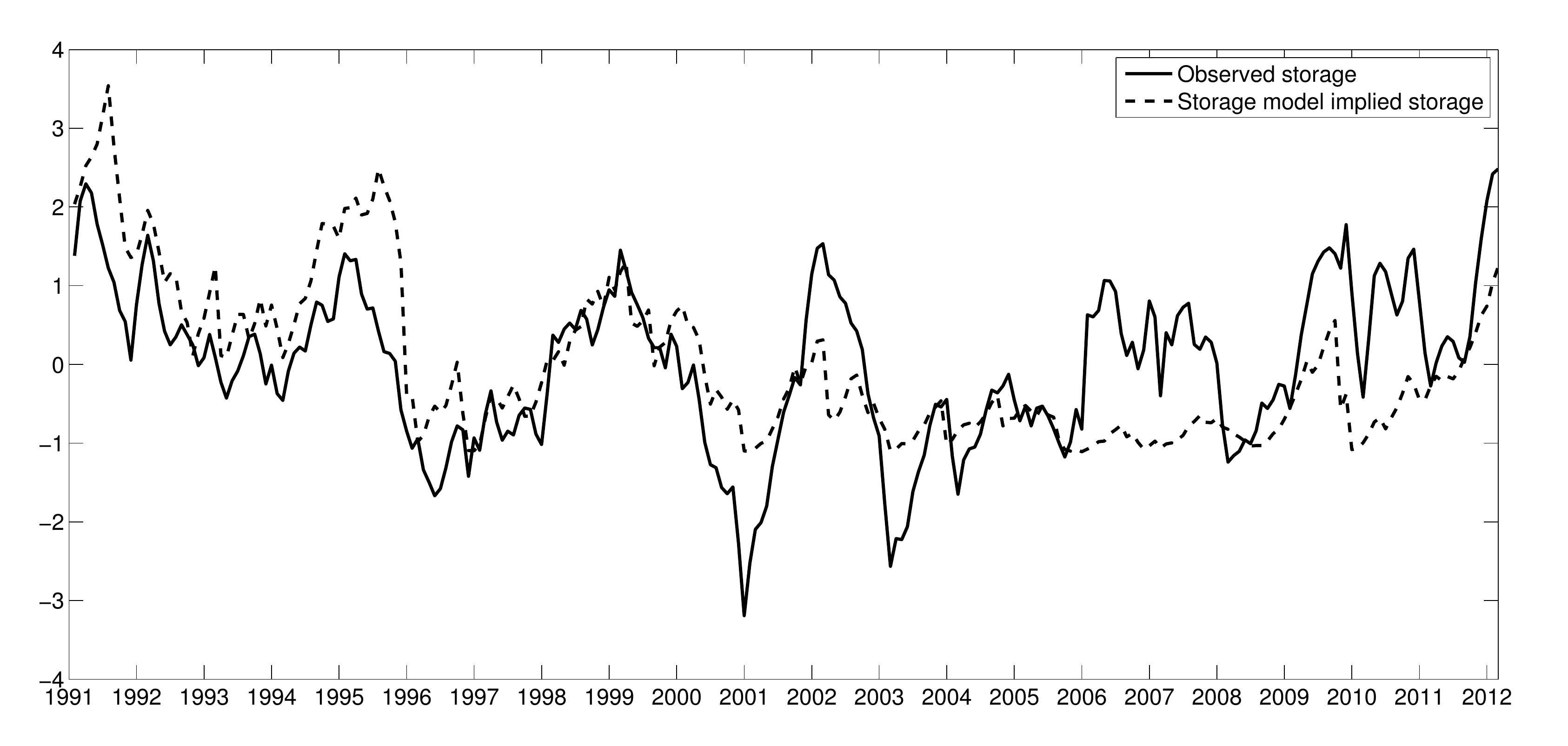}
\par\end{centering}
{\footnotesize{}Note: Observed storage is the deseasonalized and detrended
U.S. Natural Gas Underground Storage Volume (MMcf) (http://www.eia.gov/naturalgas/).
The series is deseasonalized using trigonometric functions with annual
frequency. A linear trend is used for detrending. All series' are
normalized to have mean zero and unit standard deviation. }{\footnotesize \par}

\caption{\label{fig:Figure 1}Observed and model implied storage. }
\end{figure}

Figure \ref{fig:Figure 1} plots the deseasonalized and detrended
observed monthly storage of natural gas against the median model implied
storage. 
Considering that inference on model storage only uses price data,
the two series' are similar. There is good correspondence between
the series', especially up to the early 2000's. In later years, the
variation in storage is larger than what is predicted by the storage
model. The relatively strong similarity between the series' is reassuring
as it suggests that the latent model storage is related to actual
storage. The figure gives support for the relevance of the storage
model and the importance of inventories in accounting for natural
gas price movements.
\begin{figure}
\begin{centering}
\includegraphics[scale=0.5]{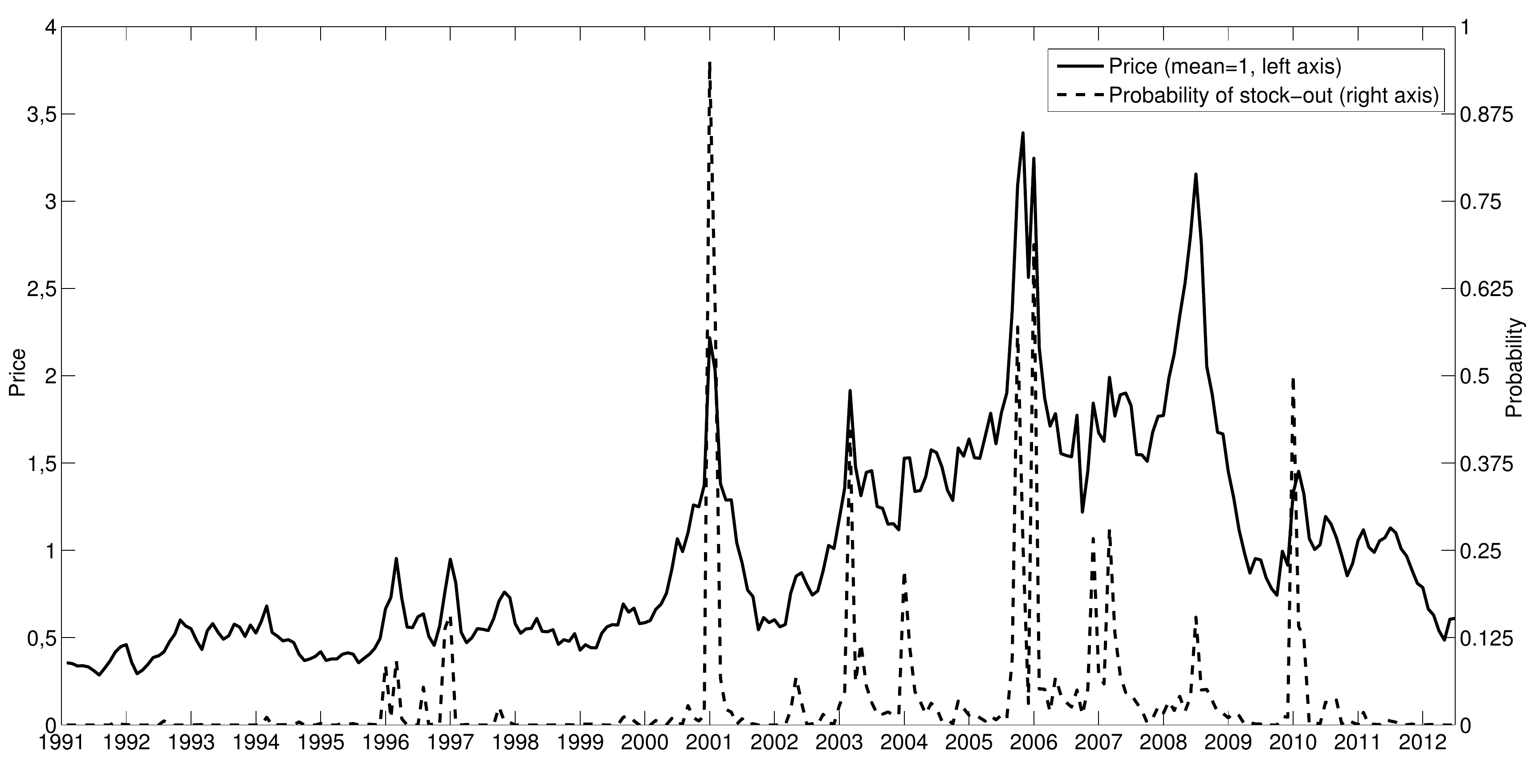}
\par\end{centering}
{\footnotesize{}Note: The filtered probabilities of stock-out at time
$t$ are calculated as $Prob(f(x_{t},z_{t})=P(x_{t})|p_{1},\dots,p_{t})$
using particle filter output. }{\footnotesize \par}

\caption{\label{fig:Figure 2}Natural gas price and model suggested probability
of stock-out }
\end{figure}

Finally, to explore the effect of the non-negativity constraint on
price dynamics we plot the price series (unit mean) and the filtered
model implied probability of a stock-out. This is shown in Figure
\ref{fig:Figure 2}. The probability of a stock-out is the probability
that the next period stock falls below a level where any positive
speculative storage is optimal. The constraint is the source of non-linearity
in the model, and as the figure shows the constraint is likely to
be binding in certain periods. These are periods associated with abnormally
high prices. Looking at the storage data in Figure \ref{fig:Figure 1}
we observe that periods with higher probability of a stock-out coincide
with periods of low storage. The figure highlights the importance
and relevance of the non-negativity constraint in accounting for characteristics
in the price data.

\subsection{Diagnostics\label{subsec:Diagnostics} }

\begin{table}
\begin{centering}
\begin{tabular}{lccc}
\hline 
 & \multicolumn{3}{c}{Natural gas data}\tabularnewline
\cline{2-4} 
 & Data & Storage model & AR(1)\tabularnewline
\hline 
$E(p_{t})$ &       1 & 0.86 & 1.02\tabularnewline
$SD(p_{t})$ & 0.61 &  0.67 &  0.61\tabularnewline
$SK(p_{t})$ & 1.29 & 2.26 &  -0.05\tabularnewline
$EKU(p_{t})$ & 1.68 & 6.46 & 0.03\tabularnewline
$AC_{1}(p_{t})$ &       0.95 & 0.96 & 0.95\tabularnewline
$AC_{2}(p_{t})$ &       0.90 & 0.94 & 0.90\tabularnewline
$AC_{1}(|\Delta p_{t}|)$ &  0.47 &  0.40 & 0.00\tabularnewline
\hline 
\end{tabular}
\par\end{centering}
\begin{raggedright}
{\footnotesize{}Note: The statistics for the estimated models are
calculated over simulated realizations of length 100,000. $SD$ denotes
standard deviations, $SK$ skewness, $EKU$ excess kurtosis, $AC_{k}$
the $k$-th order autocorrelation and in particular $AC_{1}(|\Delta p_{t}|)$
denotes the first order autocorrelation of the absolute price returns
as a measure of volatility clustering.}\caption{\label{tab:Comparison-of-statistics}Comparison of statistics from
data and estimated models. }
\par\end{raggedright}
\end{table}
As a first set of diagnostics, we report in table \ref{tab:Comparison-of-statistics}
several statistics of the actual natural gas price along with the
corresponding statistics for long realizations of the fitted storage
and linear AR(1) models. It is seen that the AR(1) model captures
better the mean, standard deviation and autocorrelation of the price
series, whereas the storage model does a better job at capturing skewness,
excess kurtosis and the autocorrelation of absolute price returns.
The latter is used as a measure of volatility clustering, and it is
seen that the storage model matches the data rather well in this sense.
The large standard deviation of the storage model for the salmon data
stems from the fact that this model produces infrequent price spikes
that are substantially larger than what is seen in the data.

As a further diagnostic of model fits, we perform a battery of tests
on the residuals. The results are given in Table \ref{tab:Diagnostics-for-residuals}.
\begin{table}
\begin{centering}
\begin{tabular}{lcc}
\hline 
 & \multicolumn{2}{c}{Residuals from }\tabularnewline
 & \multicolumn{2}{c}{Natural Gas Data}\tabularnewline
\cline{2-3} 
 & storage  & AR(1) \tabularnewline
 & model & model\tabularnewline
\hline 
Mean & 0.0175 &  -0.0001\tabularnewline
Standard Deviation & 0.9742 &  1.0001\tabularnewline
Skewness & 0.5999 &  -0.2546\tabularnewline
Excess Kurtosis & 0.5668 &  7.2735\tabularnewline
First Order Autocorrelation & 0.1877 &  0.0252\tabularnewline
Jarque-Bera $p$-value & 0.0026 & <0.0010\tabularnewline
Kolmogorv-Smirnoff $p$-value & 0.3493 & <0.0001\tabularnewline
Ljung-Box test (lag=20) $p$-value & 0.0014 & 0.0143\tabularnewline
Engle ARCH test $p$-value & 0.7648 & <0.0001\tabularnewline
\hline 
\end{tabular}
\par\end{centering}
\raggedright{}{\small{}Note: The residuals for the storage model are
calculated by a standard Gaussian quantile transform applied to the
generalized residuals (\ref{eq:generalized_residuals}). We expect
that the residuals are approximately iid standard Gaussian if the
model specification is correct.} \caption{\label{tab:Diagnostics-for-residuals}Diagnostics for residuals of
the real data set. }
\end{table}
This diagnostics also helps highlight what features of the data the
storage model addresses. Again, it is seen that the storage model
in general does a better job accounting for higher moment characteristics.
The storage model predicts non-linear first-order Markov prices. These
features will manifest in the higher moments of prices, and it is
not surprising that these are the characteristics the storage model
best describes. Not surprising, the AR(1) benchmark does a better
job when it comes to the unconditional mean price and first order
autocorrelation, while the storage model does better in terms of excess
kurtosis. In addition, the storage model better accounts for ARCH
effects. In terms of the Gaussian structure of our parametric specification
(equations (\ref{eq:stoch_model1}-\ref{eq:stoch_model3})), the diagnostic
results indicate that residuals are close to normal. 

To assess whether we are using a sufficiently fine grid, we estimate
the model on the natural gas and salmon data using different values
of $M_{x,1}$, $M_{x,2}$ and $M_{z}$. As was demonstrated by \citep{cafiero2011empirical}
under the $\rho=0$ model, using too coarse grids can bias parameter
estimates. Parameter estimates obtained while keeping the random number
seed in the particle filter fixed are provided in Table \ref{tab:diff_grid}.
The settings of the lowermost rows in each panel of the table are
the ones used for the above estimation on natural gas and salmon prices.
For the natural gas data, differences in parameter estimates for finer
grids are minor comparing to statistical standard errors from the
SML panel in Table \ref{tab:Parameter-estimates-natgas}. All in all
we conclude that we are using sufficiently fine grids for these ranges
of parameters.
\begin{table}
\begin{centering}
\begin{tabular}{cccccccc}
\hline 
$M_{x,1}$ & $M_{x,2}$ & $M_{z}$ & $\rho$ & $a$ & $b$ & $\delta$ & log-likelihood\tabularnewline
\hline 
256 & 256 & 128 & 0.9670 & 1.5257 & -0.4028 &  0.0217  & 194.2100\tabularnewline
128 & 128 & 128 & 0.9672 & 1.5106 & -0.4103 &  0.0217 & 194.3057\tabularnewline
256 & 256 & 64 & 0.9671 & 1.4781 & -0.3994 &  0.0211 & 194.2546\tabularnewline
128 & 128 & 64 & 0.9666 & 1.4819 & -0.4003 &  0.0217 & 194.3273\tabularnewline
\hline 
\end{tabular}\\
{\footnotesize{}Note: The figures in the table were obtained by estimating
the parameters using the real data sets. The random number seed was
kept fixed, and it is seen that errors associated with price function
solver discretization are small relative to the statistical standard
deviations.}
\par\end{centering}{\footnotesize \par}
\centering{}\caption{\label{tab:diff_grid}Maximum likelihood estimates and optimal log-likelihoods
for different resolutions of the rational expectation solver grids
for the real data. }
\end{table}

\section{Conclusion}

We propose a particle filter estimator for the competitive storage
model with temporal supply shock dependence when only price data is
available for estimation. The particle filter estimator utilizes information
in the conditional distribution of prices when temporal dependence
is present in the shocks. This is contrary to the composite pseudo
maximum likelihood estimator of \citet{Deaton1995,Deaton1996}, which
only utilizes the marginal state distribution to arrive at predictive
price moments. To our knowledge this is the first attempt at using
particle filter methods to estimate the competitive storage model.
Our results suggest that the relative simplicity and low-dimensional
nature of the partial equilibrium storage model makes it particularly
suitable for particle filter estimators. 

We demonstrate through simulation experiments that our particle filter
estimator does a better job in terms of both the bias and precision
of the structural parameter estimates compared to the composite estimator.
Furthermore, our estimator is less computationally demanding than
our Monte Carlo based implementation of the Deaton and Laroque composite
maximum likelihood estimator, and also is more numerically stable.
In addition, simulation evidence indicates that our model has favorable
large sample properties where bias diminishes when more price data
becomes available. The composite maximum likelihood estimator does
not display this same general reduction in bias as sample size increases. 

As an application and demonstration of the estimator we estimate the
storage model to natural gas prices. As a benchmark, we compare the
storage model to a linear AR(1) model, a GARCH(1,1) model and a two-state
Markov Switching AR(1) model. The linear AR(1) model can be thought
of as the price representation that would occur with zero speculative
storage in the storage model. The comparison to reduced form models
allows us to identify what features of prices the storage model is
able to account for, and where improvements are needed. For the natural
gas market, the storage model performs better than all reduced form
time-series models considered. This suggests that the non-linearity
in the model, arising from the non-negativity constraint on storage,
is relevant to account for natural gas price characteristics. As support
for the relevance of the storage model in the natural gas market,
we find relatively strong correspondence between observed and model
implied storage. Diagnostics show that the storage model addresses
features in the higher moments of prices, specifically linked to excess
kurtosis and ARCH effects. 

The particle filter estimator appears superior to the composite maximum
likelihood estimator for the type of model setting analyzed in this
paper. The storage model is used to investigate effects of various
commodity market policies. The value of such policy evaluations depend
crucially on the validity of the structural parameters chosen for
the analysis. When price data is the only reliable data available
to infer structural parameters, and shocks are suspected to have temporal
dependence, the particle filter estimator should be applied to efficiently
utilize the sparse data.

\singlespacing\bibliographystyle{chicago}
\bibliography{kleppe1}
\doublespacing

\appendix

\section{Fourier transform-based continuous resampling routine}

Suppose we wish to sample from a univariate Gaussian mixture on the
form
\[
\pi(z)=\sum_{j=1}^{N}w^{(j)}\mathcal{N}(z|\mu^{(j)},\sigma^{2}).
\]
Then the practical fast Fourier transform routine consist of the following
steps:
\begin{itemize}
\item Grid: Find the mean and standard deviation of $\pi(z)$ and initiate
a $n_{g}$-point regular grid containing say the mean $\pm$ 8 standard
deviations. We set $n_{g}$ to 1024 in all computations presented
in this paper.
\item PDF: As the variance in each component of $\pi(z)$ is equal, the
PDF may be computed using fast Fourier transform methods on the regular
grid as explained thoroughly in \citet{Si86}, section 3.5 (with the
modification that each particle weight is now $w^{(j)}$ and not $1/n$).
\item CDF: Compute the cumulative distribution function (CDF) of the approximate
probability density function on the same grid using a mid-point rule
for each grid point.
\item Fast inversion: Sample approximate random variables from $\pi(z)$
based on stratified uniforms using the CDF-inversion algorithm provided
in Appendix A.3 of \citet{Malik2011}.
\end{itemize}
The total operation count of this algorithm is $O(N+n_{g}\log_{2}(n_{g}))$
and thus is it linear in complexity in the number of particles retained
also for this form of continuous sampling with fixed $n_{g}$. However,
it is worth noticing that to obtain the asymptotically correct random
draws (i.e. exact quantiles corresponding to the stratified uniform
random numbers) as $N\rightarrow\infty$, $n_{g}$ must also grow,
e.g. as $O(N)$. Moreover, the area covered by the grid must also
grow, but at a slower rate, e.g. $O(\log(N))$.
\end{document}